\documentclass[prl,twocolumn,floatfix,superscriptaddress,citeautoscript,nobibnotes]{revtex4}

\usepackage{graphicx}
\usepackage{dcolumn}   
\usepackage{bm}        
\usepackage{amssymb}   
\usepackage{verbatim}
\usepackage{multirow}
\usepackage{xcolor} 
\usepackage{amsmath}
\usepackage{amsfonts}
\usepackage{soul}
\usepackage{hyperref}
\usepackage{makecell}
\hypersetup{colorlinks=true, pdfstartview=FitV, linkcolor=blue, citecolor=black, plainpages=false, pdfpagelabels=true, urlcolor=blue}
\usepackage[all]{hypcap}

\begin{document}
\title{Phonon downconversion to suppress correlated errors in superconducting qubits}

\author{V. Iaia}
\thanks{These authors contributed equally}

\affiliation{Department of Physics, Syracuse University, Syracuse, New York 13244-1130}
\author{J. Ku}
\thanks{These authors contributed equally}

\affiliation{Department of Physics, Syracuse University, Syracuse, New York 13244-1130}
\author{A. Ballard}

\affiliation{Department of Physics, Syracuse University, Syracuse, New York 13244-1130}
\author{C. P. Larson}

\affiliation{Department of Physics, Syracuse University, Syracuse, New York 13244-1130}
\author{E. Yelton}

\affiliation{Department of Physics, Syracuse University, Syracuse, New York 13244-1130}
\author{C.~H. Liu}

\affiliation{Department of Physics, University of Wisconsin-Madison, Madison, Wisconsin 53706}
\author{S. Patel}

\affiliation{Department of Physics, University of Wisconsin-Madison, Madison, Wisconsin 53706}
\author{R. McDermott}

\affiliation{Department of Physics, University of Wisconsin-Madison, Madison, Wisconsin 53706}
\author{B. L. T. Plourde}

\email[]{bplourde@syr.edu}
\affiliation{Department of Physics, Syracuse University, Syracuse, New York 13244-1130}

\date{\today}

\begin{abstract}
Quantum error correction can preserve quantum information in the presence of local errors, but correlated errors are fatal. For superconducting qubits, high-energy particle impacts from background radioactivity produce energetic phonons that travel throughout the substrate and create excitations above the superconducting ground state, known as quasiparticles, which can poison all qubits on the chip. We use normal metal reservoirs on the chip back side to downconvert phonons to low energies where they can no longer poison qubits. We introduce a pump-probe scheme involving controlled injection of pair-breaking phonons into the qubit chips. We examine quasiparticle poisoning on chips with and without back-side metallization and demonstrate a reduction in the flux of pair-breaking phonons by over a factor of 20. We use a Ramsey interferometer scheme to simultaneously monitor quasiparticle parity on three qubits for each chip and observe a two-order of magnitude reduction in correlated poisoning due to background radiation.

\end{abstract}

\maketitle

Qubits formed from superconducting integrated circuits are one of the leading systems for implementation of a fault-tolerant quantum computer \cite{Kjaergaard2020}. 
For sufficiently high gate fidelity, error correction schemes such as the surface code \cite{Fowler2012} can mitigate local errors. However, recent work has shown that high-energy particle impacts from low-level radioactivity and cosmic-ray muons will generate nonequilibrium quasiparticles (QPs) \cite{Cardani2021,Catelani2011,Vepsalainen2020} that can lead to correlated errors across a multiqubit array \cite{Wilen2021, Martinis2021,McEwen2021}. Such correlated errors cannot be mitigated by current error correction schemes, thus posing a significant challenge to realization of a fault-tolerant quantum computer.

Particle impacts deposit energy of order 100~keV in the device substrate, leading to the generation of large numbers of electron-hole pairs and a cascade of high-energy phonons \cite{Wilen2021}. These phonons travel throughout the chip and break Cooper pairs with high probability when they scatter off superconducting structures on the device layer, thus generating QPs at arbitrary locations relative to the particle impact site \cite{Martinez2019, Martinis2021,Wilen2021}. Prior work has explored low-gap superconducting structures for phonon downconversion to protect superconducting resonators and detectors with a higher gap energy \cite{Henriques2019, Karatsu2019}. Another scheme involves placing superconducting detectors on thin suspended membranes \cite{deVisser2021}. The use of normal metal layers on the back side of superconducting qubit chips was proposed in Ref.~\cite{Martinis2021} to downconvert energetic phonons below the superconducting gap. Because this downconversion process is based on the scattering of phonons with conduction electrons in the metal, the low rate of electron-phonon scattering at low temperatures \cite{Wellstood1994} dictates large volumes of normal metal for efficient phonon downconversion. A calculation in Ref.~\cite{Martinis2021} based on the phonon scattering rate in the normal metal on the back side and the pair-breaking rate in the superconducting film on the device layer indicates that achieving a 100-fold improvement in the qubit energy relaxation time $T_1$ in the aftermath of a particle impact requires a 6-$\mu$m-thick normal metal layer.

Here we implement this idea using thick electroplated Cu reservoirs to promote downconversion of phonons below the superconducting gap edge. 
To test this approach in a controlled way, we integrate Josephson junctions around the chip perimeter and controllably bias the junctions above the superconducting gap to generate pair-breaking phonons on demand. 
With explicit phonon injection, we find that the phonon-downcoversion structures reduce the flux of pair-breaking phonons by more than a factor of 20.
We also examine the correlated errors in multiqubit chips with and without Cu reservoirs and find a two order of magnitude reduction in correlated error rate, to the point where these errors no longer pose a limit to fault-tolerant operation.

\begin{figure}[htbp!]
\centering
\includegraphics[width=3.35in]{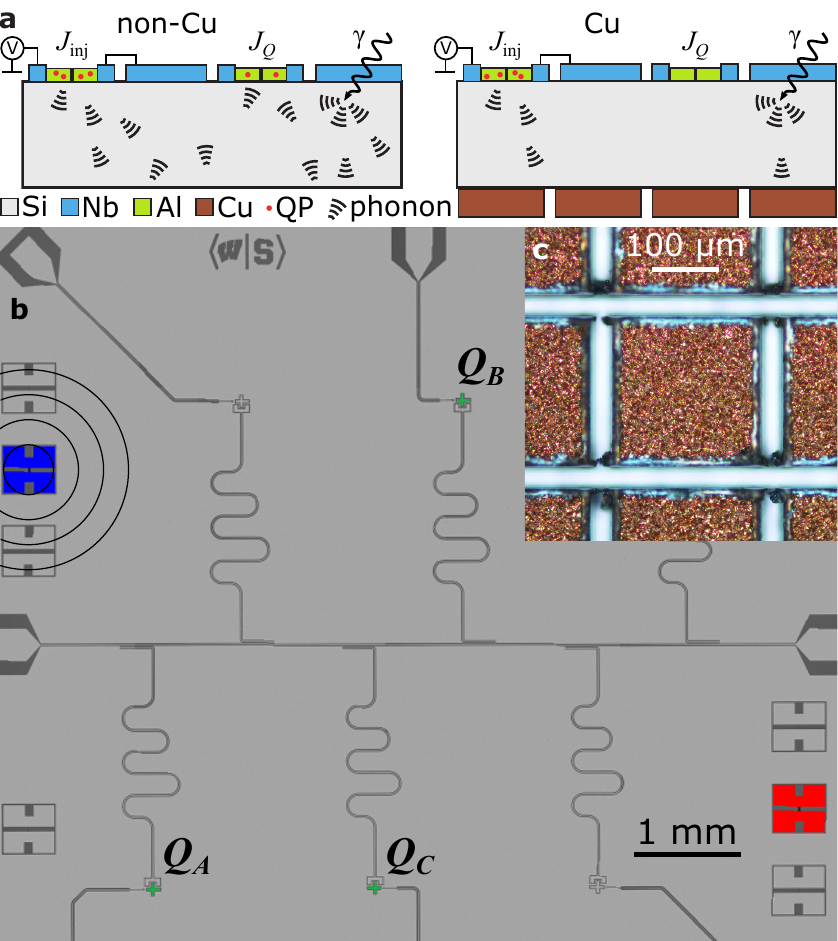}
 \caption{{\bf Phonon-mediated QP poisoning and device layout.} (a) Schematic showing QP/phonon injection, $\gamma$ impact, phonon propagation in substrate, and pair breaking in qubit junctions with and without Cu. (b) Optical micrograph of device layer. Qubits ($Q_{A,B,C}$) are colored green. Junctions used to inject QPs into the Cu (non-Cu) chip are highlighted in blue (red); concentric rings represent propagating phonons.
 (c) Optical micrograph of Cu islands on back side of Cu chip.
\label{fig:setup}}
\end{figure}

\vspace{4mm}
\noindent\textbf{\large Results}

\noindent\textbf{Experimental design.} The experimental geometry is shown in Fig.~\ref{fig:setup}. We study two nominally identical chips, one with back-side normal metallization (Cu chip) and one without (non-Cu chip). Each chip incorporates an array of charge-sensitive transmon qubits, with Josephson injector junctions arrayed around the perimeter of the chip. Each qubit has a readout resonator that is inductively coupled to a common feedline that can be used for multiplexed readout. We measure both chips in the same low-temperature environment on the same cooldown. For our qubits, we target a somewhat low ratio $E_{\rm J}/E_{\rm c}$ of Josephson energy to single-electron charging energy to produce a peak-to-peak charge dispersion between 1-5~MHz. 
This allows us to monitor QP parity switching for each qubit, which is a sensitive measure of QP poisoning \cite{Riste2013,Serniak2018,Serniak2019}.
For the experiments presented here, we focus on three of the qubits on each chip: $Q_A$, $Q_B$, $Q_C$ [Fig.~\ref{fig:setup}(b)]. Details of the qubit parameters and experimental configuration are given in Methods and Supplementary Note~3-4.

The normal metal reservoirs on the Cu chip consist of 10-$\mu{\rm m}$ thick islands patterned from Cu films grown by electrodeposition onto the back side of a high-resistivity double-side polished Si wafer following electron-beam evaporation of a Ti/Cu seed layer. The 10-$\mu{\rm m}$ thickness was chosen based on the estimate from Ref.~\cite{Martinis2021}. The islands are defined with a lattice of partial
dicing saw cuts through the Cu film into the back side of the wafer, resulting in island areas of $(200\,\mu{\rm m})^2$ [Fig.~\ref{fig:setup}(c)] (see Methods and Supplementary Note~1-2). This is done to suppress damping from coupling to the transmission line mode formed by a continuous metal layer on the back side of the chip and the ground plane on the device layer that would otherwise degrade qubit coherence \cite{Martinis2021}. Metallic losses due to capacitive coupling between the qubit and Cu island are projected to limit the qubit quality factor to $\sim3$M for the qubit design considered in Ref.~\cite{Martinis2021}; the smaller qubit island size for our qubits reduces this coupling and raises this quality factor limit, thus making a negligible impact on $T_1$.

The injector junctions are fabricated at the same time as the qubit junctions with a standard Al-AlO$_x$-Al process. The ground plane, qubit capacitor islands, readout resonators, and injector junction pads are all fabricated from Nb. There is no direct galvanic connection from the injector junction pads to ground, so the QP poisoning proceeds via phonon emission. By biasing the injector junction above $2\Delta_{\rm Al}/e$, where $\Delta_{\rm Al}$ is the superconducting energy gap for Al, we break Cooper pairs and generate local QPs that subsequently recombine, emitting phonons. These phonons then travel through the Si, and, upon encountering Al junction electrodes for a qubit, a phonon will break a Cooper pair with high probability and generate two QPs. 
A similar injection scheme was used in Ref.~\cite{Patel2017,Mannila2021}. Although the phonons injected by the tunnel junction will be lower in energy than those generated by a particle impact, the tunnel junction gives us the ability to control the timing, duration, and location of the phonon injection, in contrast to phonons from particle impacts, which occur at random times and locations.

\vspace{2mm}
\noindent\textbf{Enhanced relaxation from phonon injection.} In a first series of experiments, we measure the energy relaxation time $T_1$ of all three qubits on each chip following pulsed QP injection. Here we focus on $Q_C$ ($Q_B$) for the non-Cu (Cu) chip, but we observe similar behavior for the other two qubits on each chip. 
We use a standard inversion recovery measurement to probe $T_1$. To quantify degradation in $T_1$, we plot $\Delta \Gamma_1 = 1/T_1-1/T_1^{b}$, where $T_1^{b}$ is the baseline relaxation time from an average of several $T_1$ measurements with no injector junction bias (see Methods). The change $\Delta x_{\rm qp}$ in reduced QP density in the qubit junction leads can be calculated from $\Delta\Gamma_1$ \cite{Wang2014} (see Methods). 

\begin{figure}[ht!]
\centering
\includegraphics[width=3.4in]{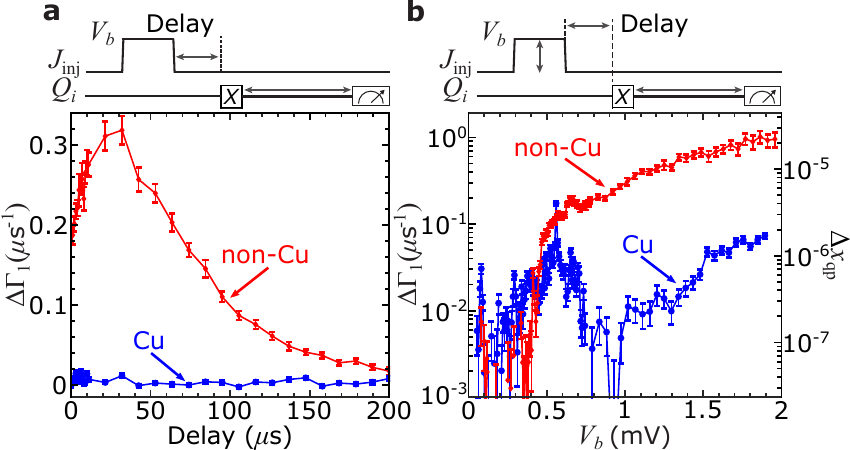}
  \caption{{\bf Suppression of $T_1$ from controlled phonon injection.} 
  (a) $\Delta\Gamma_1$ vs. delay following injection pulse for $Q_C$ on non-Cu (red) chip and $Q_B$ on Cu (blue) chip with $V_b=1$~mV. (b) $\Delta\Gamma_1$ vs. $V_b$ for non-Cu and Cu chips with 30-$\mu$s delay. Error bars computed from 95\% confidence intervals from $T_1$ fits (see Supplementary Note 5).
\label{fig:delta-gamma}}
\end{figure}

We start by applying a 10-$\mu{\rm s}$ injection pulse with amplitude $V_b$ = 1~mV, well beyond $2\Delta_{\rm Al}/e$, so that we expect significant QP poisoning in the absence of any mitigation. We vary the delay time between the end of the injection pulse and the $X$ pulse at the start of the $T_1$ sequence. For the non-Cu chip, $\Delta\Gamma_1$ increases substantially following the injection pulse and reaches a maximum poisoning level about $30\,\mu{\rm s}$ after the end of the injection pulse [Fig.~\ref{fig:delta-gamma}(a)]. This delayed onset of poisoning is consistent with the propagation timescale for the injected phonons to diffuse through the substrate to the qubit junction. In the absence of phonon downconversion structures, phonons travel throughout the substrate following boundary-limited diffusion, where they scatter randomly off the top and bottom surfaces of the Si chip.
This leads to an effective diffusivity $D=c_s d$, where $c_s=6\times 10^3\,{\rm m/s}$ is the speed of sound in Si and $d=0.525\,{\rm mm}$ is the chip thickness. Thus, the timescale for phonons to diffuse from the injector junction to each qubit (at a separation of 4-6~mm from the injector) is of the order of $10\,\mu{\rm s}$. Following this peak, $\Delta\Gamma_1$ recovers towards the unpoisoned baseline level following an exponential decay
with a characteristic time of $\sim$60~$\mu{\rm s}$ (see Supplementary Note~5). This corresponds to the timescale for phonons to exit the substrate at the chip perimeter where the sample is acoustically anchored to the device enclosure. The chip is attached to the machined Al enclosure using a small amount of low-temperature adhesive (GE varnish) at the corners of the chip.

For the Cu chip, $\Delta\Gamma_1$ is difficult to distinguish from the baseline level for all delays. For a 30-$\mu$s delay, $\Delta\Gamma_1$ for the non-Cu chip is over a factor of 35 larger than for the Cu chip. This is our first key result demonstrating the effectiveness of Cu reservoirs in reducing phonon-mediated QP poisoning.

We then explore the variation of $\Delta\Gamma_1$ with $V_b$ for a fixed delay of $30\,\mu{\rm s}$ and a 10-$\mu$s pulse width. For the non-Cu chip, we observe a significant increase in $\Delta\Gamma_1$ when the pulse amplitude exceeds $2\Delta_{\rm Al}/e\approx$~0.4~mV. For the Cu chip, $\Delta \Gamma_1$ doesn't change significantly at $2\Delta_{\rm Al}/e$; however, we observe a gradual rise in $\Delta \Gamma_1$ to a peak at a pulse amplitude of 0.56~mV, followed by a reduction to the baseline level for larger pulse amplitudes. We understand the peak in $\Delta\Gamma_1$, which is also visible for the non-Cu chip on top of the overall poisoning curve, to be due to photon-assisted poisoning from absorption of Josephson radiation emitted by the injector junction mediated by a spurious mm-wave resonance in the qubit (see Supplementary Note~5-6). 
Such antenna effects in qubit structures can lead to resonant absorption of electromagnetic radiation, which can drive high-frequency currents through the qubit junction and generate QPs \cite{Houzet2019,Rafferty2021,liu2022}. We would not expect the Cu reservoirs to have any effect on this photon-based QP poisoning mechanism. 

\begin{figure*}[!ht]
\centering
\includegraphics[width=6.8in]{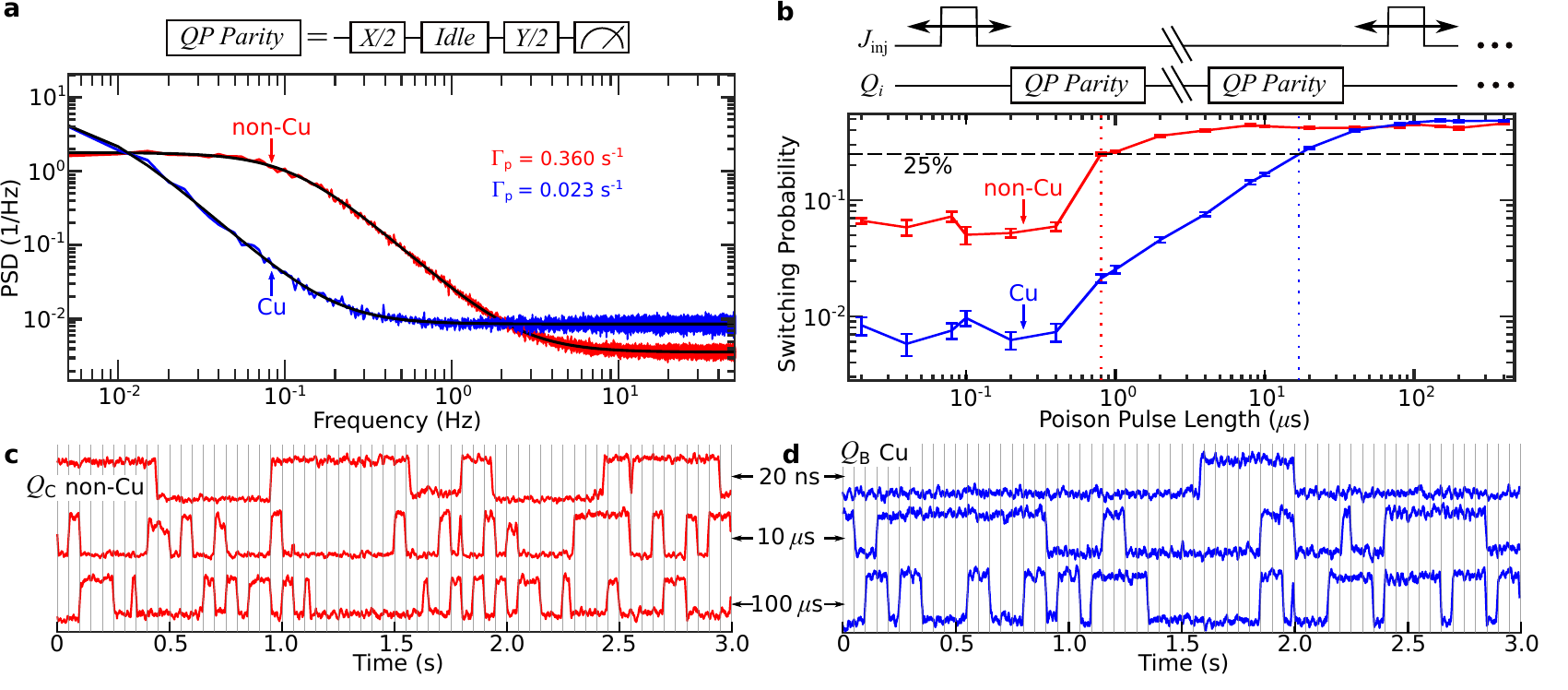}
    \caption{{\bf Measurement of QP parity switching.}
    (a) Power spectral density of QP parity switching with no injection pulses for $Q_A$ on non-Cu (red) and Cu (blue) chips. (b) Measured probability of parity switch per injection pulse vs. pulse duration for non-Cu and Cu chips; dotted/dashed lines indicate pulse lengths corresponding to 25\% switching probability. Error bars computed from standard Poisson counting errors (see Supplementary Note 9). Pulse sequence for QP parity measurements without/with controlled phonon injection shown above plots in (a/b). Segment of time series of QP parity for different injection pulse durations for (c) non-Cu chip, (d) Cu chip; vertical lines indicate timing of injection pulses. 
\label{fig:poisoned-parity}}
\end{figure*}

\vspace{2mm}
\noindent\textbf{QP parity switching.} In a separate series of experiments, we exploit the non-negligible charge dispersion of our qubits to probe the charge parity of the qubit islands as a sensitive probe of QP poisoning. We employ a Ramsey pulse sequence to map QP parity onto qubit 1-state occupation \cite{Riste2013, Serniak2018, Christensen2019} (see Methods). We first perform the QP parity switching measurement on each of the qubits at a repetition period of 10~ms and then compute the power spectral density of parity switches. We fit a Lorentzian to the measured spectrum for each chip to extract the characteristic parity switching rate $\Gamma_{\rm p}$, as in Ref.~\cite{Riste2013}. In Fig.~\ref{fig:poisoned-parity}(a), we plot typical spectra from one qubit on each chip. The resulting values for $\Gamma_{\rm p}$ for both chips are low: $\Gamma_{\rm p}$ = 0.360~s$^{-1}$ (0.023~s$^{-1}$) for the non-Cu (Cu) chip. To the best of our knowledge, $\Gamma_{\rm p}$ for the non-Cu chip is consistent with the lowest rates for QP poisoning reported in the literature \cite{Pan2022,Kurter2021}, while for the Cu chip our measured poisoning rate is an order of magnitude lower. 

The low QP poisoning rates on both chips are likely due to a combination of best practices for shielding, filtering, and thermalization (see Supplementary Note~3). In addition, the relatively compact qubit design results in a rather high fundamental antenna resonance frequency, $\sim$270~GHz (see Supplementary Note~6), which is likely above the cutoff of the spectrum of blackbody radiation from higher temperature stages of the cryostat. Following the analysis in Ref.~\cite{liu2022} applied to the geometry of our qubits, we calculate an effective blackbody temperature of 330~mK (280~mK) for $Q_A$ on the non-Cu (Cu) chip from our measured $\Gamma_{\rm p}$ values. In addition to photon absorption by the spurious antenna resonance, the residual QP poisoning is likely due to high-energy particle impacts or other radiation sources that generate pair-breaking phonons. We attribute the even lower QP parity switching rate for the Cu chip to absorption of a significant fraction of the phonons generated from these poisoning events by the Cu islands on the back side of the chip, which we will subsequently quantify. 

The low QP baseline poisoning rates allow us to directly investigate QP poisoning in the presence of controlled injection of pair-breaking phonons into the chip. Here, we sample QP parity on the qubits with a 100-$\mu{\rm s}$ repetition period, while pulsing the injector junction at an amplitude of 1~mV and a fixed rate of 20~Hz. Since this experimental duty cycle is much faster than our background switching rate, we can apply a moving average over 100 time steps for both the non-Cu and Cu chip to improve the signal-to-noise ratio for these time traces. We then perform a hidden Markov model analysis to identify the parity switches (see Methods).

As we vary the pulse length from 20~ns to $400\,\mu{\rm s}$, we increase the injected energy and thus the number of pair-breaking phonons coupled to the chip. Longer injection pulses result in a higher rate of parity switches, with almost all of the parity switches synced with the phonon injection pulses [Fig.~\ref{fig:poisoned-parity}(c,d)]. In Fig.~\ref{fig:poisoned-parity}(b) we plot the ratio of the measured switching rate to the rate of phonon injection as a function of the injection pulse duration; this quantity corresponds to the probability of a measured parity switch per phonon injection pulse. For sufficiently high injected energy, we expect that pair-breaking phonons will randomize the parity on every qubit island on a timescale much shorter than our 100-$\mu{\rm s}$ sampling period. Because we observe a change of QP parity only for an odd number of switches, we expect our measured parity switching rate to saturate at half the injection rate for long injection pulses. 
As expected, our measured probabilities saturate around 0.5; however, the Cu chip requires roughly 20 times the injection energy to achieve the same level of poisoning as the non-Cu chip. If we assume that each injection pulse generates a number of pair-breaking phonons in the Si that is proportional to the pulse duration, this indicates that the Cu islands on the back side of the Cu chip downconvert 95\% of the injected phonons. 

\vspace{2mm}
\noindent\textbf{Multi-qubit correlated parity switching.} We next perform simultaneous parity measurement of all three qubits on each chip and we analyze the resulting time series to identify coincidences (see Supplementary Note~9-10). We first apply this approach to the measurements with periodic QP poisoning from controlled phonon injection. For sufficiently long injection pulses, we expect the probability of double coincidences to saturate at $(1/2)^2$ and for triple coincidences to saturate at $(1/2)^3$; this is indeed what we observe (see Supplementary Note~8). 

\begin{figure}[ht]
\centering
\includegraphics[width=3.4in]{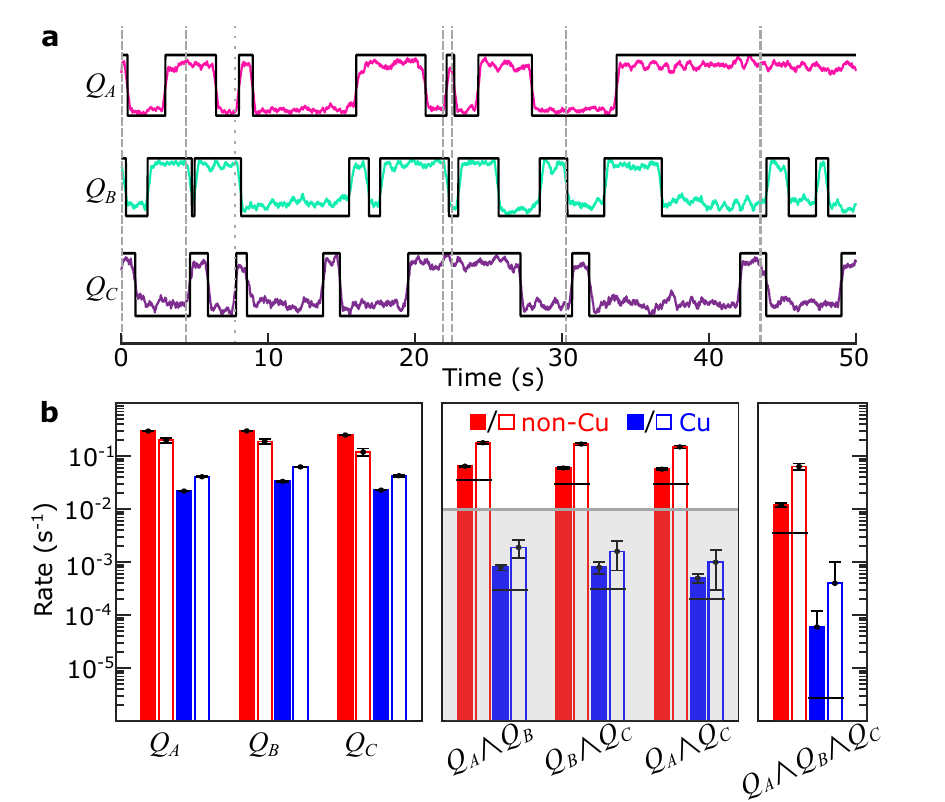}
\vspace{-15pt}
  \caption{{\bf Correlated parity switches from background radiation.} 
    (a) Typical time trace of simultaneous QP parity measurements on non-Cu chip with no injection pulses. Black lines correspond to extracted digital signal from parity switches; vertical dashed (dotted) lines indicate double (triple) coincidences. (b) Observed parity switching rates (solid bars) and extracted poisoning event rates (open bars) for non-Cu (red) and Cu (blue) chips in the absence of controlled phonon injection. Expected random background coincidence rates are plotted as horizontal black lines. Error bars computed from Poisson counting errors (see Supplementary Notes 10-11). Fault-tolerant level for two-fold correlated errors, as described in the text, is indicated by gray-shaded region.
\label{fig:unpoisoned-parity}}
\end{figure}

After confirming that our analysis successfully detects coincidences induced by controlled injection, we next apply this same approach to measurement of correlated poisoning induced by environmental radiation. Since the background poisoning rates are quite low for these devices, we reduce the experimental duty cycle to 10~ms and acquire simultaneous parity data over several days to build up sufficient statistics to detect  coincidences. Figure~\ref{fig:unpoisoned-parity} presents our results from these measurements. For all three qubits on each chip, the single-qubit parity switching rate is consistent with our previously measured $\Gamma_{\rm p}$ values. Based on the observed parity switching rates, we calculate the expected random double- and triple-coincidence rates (see Methods). For the non-Cu chip, the expected random rates for double (triple) coincidences are less than the observed coincidence rates by nearly a factor of 2 (3), indicating the presence of significant correlated switching.

Based on the analysis in Refs.~\cite{Wilen2021,Martinis2021}, in the absence of mitigation, we expect correlated poisoning events to be dominated by $\gamma$ impacts that broadcast high-energy phonons throughout the entire chip. For a given impact rate $R_{\gamma}$, we thus expect a rate of individual qubit parity switches $R_{\gamma}/2$, a rate of two-fold coincidences $R_{\gamma}/4$, and a rate of three-fold coincidences $R_{\gamma}/8$. We solve a system of equations for the observed coincidence probabilities to obtain the actual exclusive rates for single-, double-, and triple-qubit poisoning events (see Supplementary Note~11). If there were no other poisoning mechanisms and no phonon loss in the chip so that each $\gamma$ impact poisoned all qubits with 100\% probability, we would expect an extracted event rate for $Q_A\land Q_B\land Q_C$ equal to $R_{\gamma}$, while all single- and two-fold poisoning rates would be zero, since all particle impacts are expected to couple to all qubits via high-energy phonons.

Figure~\ref{fig:unpoisoned-parity}(b) presents the observed parity rates and extracted poisoning event rates for both chips. For the non-Cu chip, the extracted three-qubit event rate is high [0.064(9)~s$^{-1}$], indicating the presence of significant correlated poisoning between widely separated qubits. However, the event rates for double- and single-qubit poisoning are also significant; we note that for the three qubit pairs with different physical separations, there is no clear dependence of two-fold correlated poisoning on the distance between qubits. For any practical implementation, there will always be some degree of phonon loss, for example, from the anchoring points where the chip is attached to the sample enclosure or through wirebonds, so that even in the absence of phonon downconversion structures, not all qubits are poisoned by each $\gamma$ impact. In this case, $R_{\gamma}$ could be estimated as 1.1~s$^{-1}$, the sum of all the poisoning rates for the non-Cu chip in Fig.~\ref{fig:unpoisoned-parity}(b). For the Cu chip, all of the extracted correlated event rates are two orders of magnitude lower than for the non-Cu chip; the sum of all poisoning rates on the Cu chip is 0.15~s$^{-1}$, which is dominated by the single-qubit poisoning rates. This indicates that the Cu reservoirs greatly reduce the footprint of the phonon burst following a high-energy particle impact.

\vspace{2mm}
\noindent\textbf{\large Discussion}

We have separately performed repeated charge tomography for one qubit on each chip and observed a 
rate of large offset charge jumps 
of 0.0012(1)~[0.0011(1)]~s$^{-1}$ for the non-Cu [Cu] chip. Following the detailed modeling and analysis in Ref.~\cite{Wilen2021}, we estimate the rate of $\gamma$ impacts on our chips to be 0.083(8)~s$^{-1}$ (see Supplementary Note~13). 
Thus, the higher total poisoning rate, particularly on the non-Cu chip, compared to the estimated $R_{\gamma}$ from the offset-charge measurements suggests the presence of additional phonon-mediated poisoning mechanisms in our device. THz photons above $2\Delta_{\rm Nb}$ from warmer portions of the cryostat could break pairs in the Nb ground plane and couple phonons into the substrate from recombination, thus poisoning nearby qubits, but without the chip-wide burst of phonons from a high-energy $\gamma$ impact. Additionally, the cryogenic dark matter detection community has observed heat-only events that are attributed to mechanical cracking processes in the device and sample enclosure that release stresses, typically at the attachment points \cite{Aastrom2006, Armengaud2016}; recent work reported heat-only events in superconducting transition edge sensors on a Si chip attached to a sample holder with GE varnish \cite{Anthony2022}. Such events can produce large bursts of phonons that are detected by the sensors in these experiments, but with no accompanying charge signal. The dynamics of such heat-only events will depend on the details of the device and enclosure design, but could potentially occur in our qubit chip and sample enclosure and serve as another phonon-mediated QP poisoning mechanism. The overall reduced poisoning rates on the Cu chip indicate that the normal metal structures reduce phonon-mediated poisoning from other mechanisms in our system, such as THz photons or heat-only events, as well.

Excess QPs cause both enhanced parity switching [Fig.~\ref{fig:poisoned-parity}(b)] and reduced $T_1$ (Fig.~\ref{fig:delta-gamma}), thus resulting in enhanced bit-flip errors \cite{McEwen2021}. Thus, our demonstrated suppression of correlated QP poisoning from phonon downconversion provides a strategy for reducing correlated errors in large qubit arrays. For robust error detection, we require single-qubit errors below the 10$^{-4}$ level, which will correspond to random error coincidences between pairs of qubits at the $10^{-8}$ level. Thus, any
correlated two-qubit errors must be below the 10$^{-8}$ level \cite{Wilen2021}. If we assume a surface code duty cycle of 1~MHz 
and take our largest extracted two-fold poisoning event rate of 0.002~s$^{-1}$, we find a two-fold error probability of $2\times 10^{-9}$.
Thus, our initial attempt at correlated error suppression by phonon downconversion already yields a correlated error rate below the threshold necessary for fault-tolerant operation. Further optimization, including an investigation of the dependence of the downconversion efficiency on metal film thickness and composition, and incorporation of additional mitigation strategies should guarantee the robust operation of error-corrected quantum processors in the presence of low-level pair-breaking radiation.

\vspace{4mm}
\noindent\textbf{\large Methods}

\noindent\textbf{Device fabrication.} Both the non-Cu and Cu chips are fabricated from high-resistivity ($>$10~k$\Omega$-cm) Si wafers. For the Cu chip, the wafer is double-side polished to allow for fabrication of the Cu reservoirs. Deep-UV photolithography is used to pattern the ground plane, feedline, readout resonators, qubit islands, charge-bias lines, and injector junction pads, followed by reactive ion etching of the Nb film. After the base-layer processing, the Cu reservoir fabrication on the wafer with the Cu chip is started by first preparing a protective resist layer on the device surface, then evaporating a seed layer of Ti and Cu on the opposite side. We electroplate a 10-$\mu{\rm m}$ thick film of Cu on top of the seed layer; we pattern the Cu reservoirs by dicing $(200\,\mu{\rm m})^2$ islands with partial dicing saw cuts that extend $20\,\mu{\rm m}$ into the back surface of the Si (see Supplementary Note 1). The qubit and injector junctions on both chips are Al-AlO$_x$-Al junctions made by double-angle evaporation, producing qubit frequencies in the range of 4.7-5.3~GHz (Supplementary Table 1).

\vspace{2mm}
\noindent\textbf{Measurement setup.}
Measurements on both the non-Cu and Cu chips are performed on the same dilution refrigerator cooldown running at a temperature below 15~mK. The Al sample boxes for both chips are mounted on the same cold-finger inside a single Cryoperm magnetic shield. A Radiall relay switch on the output lines of the two devices allows us to switch between measurements of one chip or the other. Supplementary Fig. 2 details the configuration of cabling, attenuation, filtering, and shielding inside the cryostat, as well as the room-temperature electronics hardware for control and readout. The inner surfaces of the Cryoperm magnetic shield and the mixing chamber shield were both coated with an infrared-absorbent layer \cite{Barends2011}. 
For the charge-biasing of the qubits, wiring limitations on our dilution refrigerator prevented us from connecting to all of the bias traces on the chips. For the non-Cu chip, charge-bias lines are connected to $Q_B$ and $Q_C$; for the Cu chip, there is only a bias connection to $Q_A$.

\vspace{2mm}
\noindent\textbf{Relaxation and injection measurements.} Phonon injection experiments are done by pulsing the bias on one of the Josephson junctions near the edge of each chip [Fig.~\ref{fig:setup} and Supplementary Fig. 1(c)] followed by a measurement of qubit $T_1$, from which we compute $\Delta\Gamma_1$ (see Supplementary Note 5). In addition to analyzing the response of $\Delta\Gamma_1$ with bias-pulse amplitude and delay between the pulse and $T_1$ measurement, we compute the change in reduced QP density, $\Delta x_{\rm qp}=\pi\Delta\Gamma_1/\sqrt{2\Delta_{\rm Al}\omega_{01}/\hbar}$ \cite{Wang2014}, where $\omega_{01}$ is the qubit transition frequency.

\vspace{2mm}
\noindent\textbf{Single-qubit parity measurements.}
The Ramsey pulse sequence that we use for mapping QP parity onto qubit 1-state occupation is as follows: apply a $X/2$ pulse, idle for a time corresponding to a quarter of a qubit precession period, then apply a $Y/2$ pulse, followed by a qubit measurement \cite{Riste2013, Serniak2018, Christensen2019}. If the offset charge is at maximal charge dispersion, the final $Y/2$ pulse projects the qubit to either the 0 or 1 state, dependent on the QP parity. In order to have an uninterrupted measurement sequence, active  stabilization of the offset charge is not performed. The power spectral densities of the QP parity switching are computed from records of 20,000 single shots of the parity-mapping pulse sequence measured at a repetition period of 10~ms. We apply a simple thresholding scheme based on the 0/1 readout calibration for each qubit to generate a digital time trace of the QP parity. We then compute the PSD of this digital signal and average 20-160 of these curves to obtain Figure ~\ref{fig:poisoned-parity}(a) for $Q_A$ on each chip and Supplementary Fig. 7 for all three qubits on both chips. Since the offset charge is not actively stabilized, when the offset charge randomly jumps near the degeneracy point, the parity readout fidelity vanishes. This results in an enhancement of the white noise floor, but the characteristic QP parity switching rate $\Gamma_{\rm p}$ can still be extracted (see Supplementary Note 7).

In addition to the PSD measurements, we also study single-qubit QP parity switching with periodic phonon injection [Fig.~\ref{fig:poisoned-parity}(b-d)]. Here, we simultaneously produce phonons by pulsing the injector junction to an amplitude of 1~mV at a frequency of 20~Hz while recording single shots of the QP parity-mapping pulse sequence at a repetition period of 100~$\mu$s for a duration of 400~s. As with the PSD measurements, we do not actively stabilize offset charge, and thus the offset charge will occasionally jump randomly to near degeneracy where the QP parity cannot be discriminated. In order to process the data, we apply a moving average of 100 time steps to the QP parity traces. The portions of the averaged parity traces where the peak-to-peak amplitude is below a threshold determined by the 0/1 readout calibration levels are masked off and not analyzed further. Next, a hidden Markov model (HMM) is used to identify the QP parity. We assign a probability for the parity signal to have an odd- or even-parity state based on Gaussian fits to the 0/1 readout calibration measurements for each qubit. The probability for the states to transition is set by the $\Gamma_{\textrm{p}}$ from the corresponding PSD for each qubit. We then use the Viterbi algorithm to fit a digital signal to the averaged QP parity data (see Supplementary Note 9).

\vspace{2mm}
\noindent\textbf{Multi-qubit parity measurements.}
For measurements of multi-qubit QP parity switching due to background radioactivity, we perform the QP parity-mapping pulse sequence for all three qubits on a chip simultaneously at a repetition period of 10~ms. We use the previously described HMM to identify QP parity switching from the time trace for each qubit using a moving average of 40 points. This results in the averaged QP parity switching events having a sloped step, with the width of each parity switch approximately equal to the number of points used in the moving average. Following the HMM extraction of digital parity switching traces, we identify a coincidence switching event between qubits to occur when the digital time traces switch within a window of 40 data points (see Supplementary Fig. 10). The coincident events are indexed with the relevant qubits involved in the switching event: $Q_A\land Q_B$, $Q_B\land Q_C$, $Q_A\land Q_C$, or $Q_A\land Q_B\land Q_C$. We restrict each switch of a given qubit to participate in only one event per coincidence type. For example, a $Q_B$ switch cannot be used for two $Q_A\land Q_B$ coincidences, but could be used for a $Q_A\land Q_B$ coincidence and a $Q_B\land Q_C$ coincidence. The switching rate for each type of coincidence event $r_i$, where $i=AB,BC,AC,ABC$, is given by $N_i/\tau_i$, where $N_i$ is the total number of events and $\tau_i$ is the total duration of unmasked data for event type $i$. Note that double coincidences between qubits $j$ and $k$ are only counted during the period when both qubits are unmasked; similarly, triple coincidences require that all three qubits are unmasked. The uncertainty in $r_i$ comes from the standard Poisson counting errors $N_i^{1/2}/\tau_i$ (see Supplementary Note 10). 

\vspace{2mm}
\noindent\textbf{Extraction of correlated poisoning rates.}
For a set of observed single-qubit parity switching rates with a particular non-zero window $\Delta t$ for identifying coincidences, one would expect a rate of random uncorrelated coincidence switching given by the product of the probabilities for observing a parity switch for each constituent qubit during the interval $\Delta t$. These expected background coincidence parity switching rates are listed in Supplementary Table 2. The error bars for these random background coincidence rates were computed by summing the fractional uncertainty for each observed rate in quadrature. While the quantities we measure in our simultaneous parity measurements are the observed parity switching rates, we would like to compute the actual poisoning event rates $r_i$ for each qubit, or group of qubits, exclusively. For example, a single poisoning event that couples to both $Q_A$ and $Q_B$ will contribute to $r_{AB}$ but will not contribute to $r_A$ or $r_B$. Based on these criteria, we use the observed parity switching rates $r_i^{\rm obs}$ to compute the probability for observing each type of parity switching event in a window interval $\Delta t$ as $p_i^{\rm obs}=r_i^{\rm obs}\Delta t$. We then derive expressions for the probability of observing each type of parity switching event in terms of the actual probability for each type of poisoning event, as listed in Supplementary Eq.~(2). We numerically solve the system of equations to obtain the actual poisoning probabilities $p_i$ and then calculate the extracted poisoning rates $r_i=p_i/\Delta t$ reported in Fig.~\ref{fig:unpoisoned-parity}(b) and Supplementary Table 2. The error bars on each actual poisoning probability are calculated by numerically computing the derivative with respect to each of the observed switching probabilities, then multiplying by the corresponding Poisson error bar for the observed switching probability and summing these together in quadrature (see Supplementary Note 11).

\vspace{2mm}
\noindent\textbf{\large Data availability}

\noindent Data used in this work is available on \href{ https://doi.org/10.5281/zenodo.7249678}{10.5281/zenodo.7114229}. Supplementary data is available upon reasonable request.

\vspace{2mm}
\noindent\textbf{\large Code availability}

\noindent Code used in this work is available upon reasonable request.

\vspace{2mm}
\noindent\textbf{\large References}



\vspace{2mm}
\noindent\textbf{\large Acknowledgements}

\noindent This work is supported by the U.S. Government under ARO grant W911NF-18-1-0106. Fabrication was performed in part at the Cornell NanoScale Facility, a member of the National Nanotechnology Coordinated Infrastructure (NNCI), which is supported by the National Science Foundation (Grant NNCI-2025233). 

\vspace{2mm}
\noindent\textbf{\large Author contributions}

\noindent V.I. and J.K. took and analysed the data. V.I. and A.B. designed and fabricated the devices. J.K., C.P.L., and E.Y. developed code for analysing parity switching data. C.H.L., S.P., and R.M. performed the CST Microwave simulations. J.K., V.I., and A.B. helped to develop the measurement and fabrication infrastructure. B.L.T.P. designed the experiment and directed data-taking and analysis. V.I., J.K., C.P.L., E.Y., R.M., and B.L.T.P. co-wrote the manuscript.

\vspace{2mm}
\noindent\textbf{\large Competing interests}

\noindent Authors B.L.T.P and R.M. declare a competing interest in the form of a utility patent application, FABRICATION OF NORMAL CONDUCTING OR LOW-GAP ISLANDS FOR DOWNCONVERSION OF PAIR-BREAKING PHONONS IN SUPERCONDUCTING QUANTUM CIRCUITS (application number 17/469,380), filed by the Wisconsin Alumni Research Foundation on September 8, 2021. B.L.T.P and R.M. are the sole inventors on the application, which is currently pending. The invention covered by the patent application concerns the technique for fabricating the normal metal structures for phonon downconversion. This article describes experimental work by our two research groups demonstrating the effectiveness of this technique for phonon downconversion. The remaining authors declare no competing interests.

\vspace{2mm}
\noindent\textbf{\large Additional Information}

\noindent\textbf{Correspondence} and requests for materials should be addressed to B.L.T.Plourde.


\widetext
\clearpage
\begin{center}
\textbf{\large Supplementary Information: Phonon downconversion to suppress correlated errors in superconducting qubits}
\end{center}

\def\thesection{\Roman{section}}








\setcounter{secnumdepth}{3}
\setcounter{equation}{0}
\setcounter{figure}{0}
\setcounter{table}{0}
\setcounter{page}{1}
\renewcommand{\theHtable}{Supplement.\thetable}
\renewcommand{\theHfigure}{Supplement.\thefigure}
\makeatletter
\renewcommand{\theequation}{\arabic{equation}}
\renewcommand{\thefigure}{\arabic{figure}}
\renewcommand{\thetable}{\arabic{table}}

\newif\if@seccntdot


\def\@seccntformat#1{%
  \csname the#1\endcsname
  \if@seccntdot .\fi
  \quad
}

\makeatother
\renewcommand{\figurename}{Supplementary Figure}
\renewcommand{\thetable}{\arabic{table}}
\renewcommand{\tablename}{Supplementary Table}
\renewcommand{\thesection}{}
\renewcommand{\sectionautorefname}{14}

\section{Supplementary Note 1: Device Fabrication}

Both the non-Cu and Cu chips are fabricated on high-resistivity ($> 10\,{\rm k}\Omega$-cm) 100-mm Si wafers; for the Cu chip, the wafer is double-side polished to facilitate the deposition and patterning of the Cu reservoirs. Initially the wafer is put through a standard RCA clean process and then submerged in a buffered-2$\%$ per vol. HF bath to remove native oxides immediately before sputter-deposition for the non-Cu (Cu) chip of a 55-nm (80-nm) thick Nb film on the top surface of each wafer. We pattern the Nb films using a deep-UV photostepper to define the ground plane, feedline, readout resonators, qubit islands, charge-bias lines, and injector junction pads followed by a dry etch using BCl$_3$, Cl$_2$, and Ar in an ICP etcher. In the case of the non-Cu devices, we proceed with the wafer to the Josephson junction definition step, while for the Cu devices, we next fabricate the Cu reservoirs.


After stripping the base layer resist with a TMAH hot strip bath, we then coat the surface with the Nb pattern using a thick photoresist layer (SPR-220-3.0) to protect the Nb during the subsequent backside processing for the Cu reservoirs. We then deposit a metal seed layer on the back side of the wafer using electron-beam evaporation of Ti (20~nm) at a deposition rate of 1\AA/s followed by Cu (100~nm) deposited at 2\AA/s. 

For the Cu reservoirs, we deposit Cu on the wafer back side with an electrodeposition process by submerging our wafer into a copper sulfate and sulfuric acid solution. We grow a 10-$\mu{\rm m}$ thick Cu film on top of the seed layers at a rate of $\sim$3.3 $\mu{\rm m/hr}$ using an alternating current deposition mode. 
A test film grown with the same parameters and patterned into a narrow strip using Kapton tape was measured to have RRR $\sim$42. The islands were defined with a lattice of partial 50-$\mu{\rm m}$-wide dicing saw cuts through the Cu film into the back side of the wafer, with the cuts extending $20\,\mu{\rm m}$ into the back surface of the Si, resulting in island areas of $(200\,\mu{\rm m})^2$. After the Cu islands are fabricated, all resist is stripped in a TMAH hot strip bath. 

For both wafers, the Josephson junctions are then defined with a conventional double-angle shadow-evaporation process using 100~keV electron-beam lithography of a PMMA/MMA bilayer resist stack. After an {\it in situ} ion mill cleaning step to remove native oxide from the Nb surface at the contact points to the junction electrodes, the junctions are formed with electron-beam evaporation of Al. The bottom (top) junction electrode is 40~(80)~nm thick.

\begin{figure}[h!]
\centering
\includegraphics[width=6.8in]{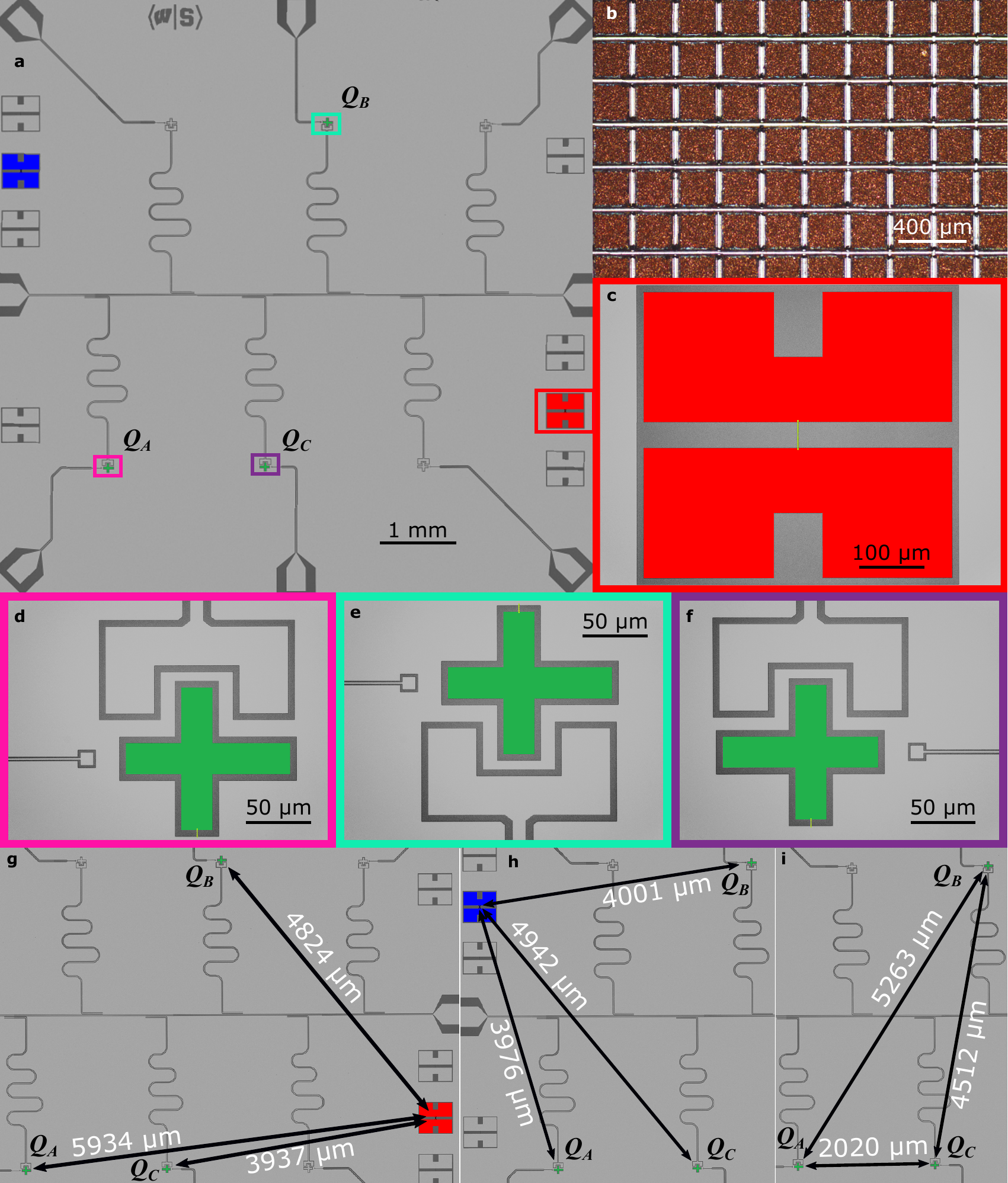}
  \caption{{\bf Optical micrographs of devices.} (a) Stitched composite image of the device layer of the chip (see Supplementary Note \autoref{sec:image_editing}\hspace*{-1mm}). (b) Cu islands on back side of the Cu chip. (c) Close-up view of the injector junction used for the non-Cu chip. Nb pads are colored in red and Al junction is highlighted yellow. (d,e,f) Close-up images of qubits ($Q_{A,B,C}$). Nb island is colored green, and Al junction electrodes are highlighted in yellow. Qubit distances from injector junction on the (g) non-Cu chip, (h) Cu chip, and (i) interqubit spacing for both chips.
  \label{fig:chipImages}}
\end{figure}
















\section{Supplementary Note 2: Device Layout}

Following the fabrication, the wafers are diced into chips that are (8~mm)$^2$. The coplanar waveguide feedline runs across the middle of the wafer, with the 1/4-wave readout resonators for each qubit inductively coupled to the feedline. A full-chip layout can be seen in Supplementary Fig.~\ref{fig:chipImages}, along with close-up views of each qubit, the non-Cu injector junction, and the Cu island pattern on the back side of the Cu chip. Editing of the individual micrographs to obtain the full-chip image is described in Supplementary Note \autoref{sec:image_editing}\hspace*{-1mm}. 
The locations of the qubits measured in the experiment relative to the injector junctions used for controlled QP poisoning, as well as the inter-qubit separations, are indicated in Supplementary Fig.~\ref{fig:chipImages}(g-i).


\begin{figure}[h!]
\centering
\includegraphics[width=7in]{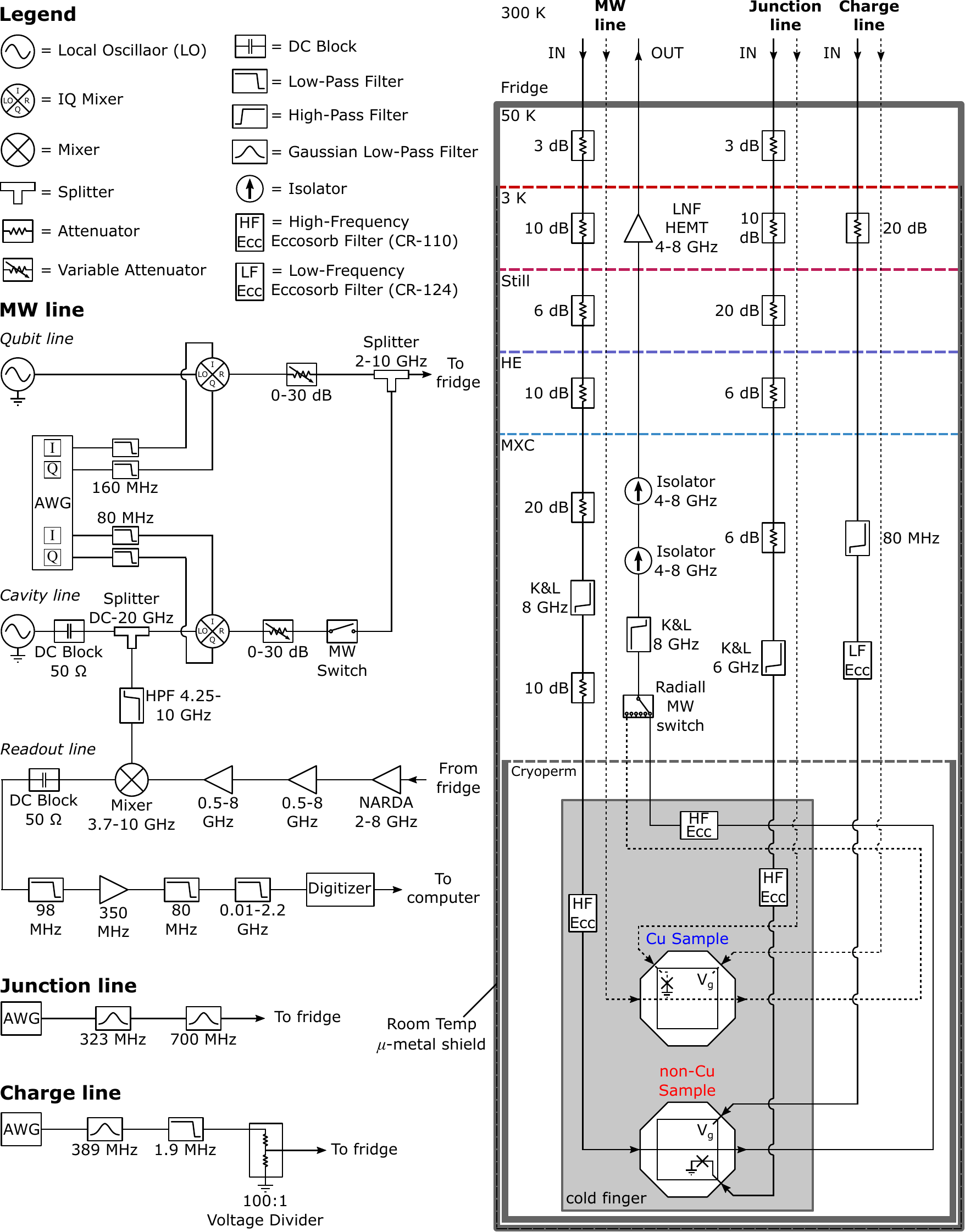}
  \caption{{\bf Experimental configuration.} Wiring diagram for room-temperature components on the left and cryogenic wiring on the right with different temperature stages indicated. Cryoperm magnetic shield and mixing chamber (MXC) shields have an IR-absorbent coating. 
  All filters are LPF unless stated otherwise. The dotted lines in the fridge represent the MW, junction, and charge lines for the Cu chip and are configured identically to the corresponding lines for the non-Cu chip.
  \label{fig:fridgeSchematic}}
\end{figure}

\section{Supplementary Note 3: Device and Measurement Setup}

Measurements on both the non-Cu and Cu chips are performed on the same dilution refrigerator cooldown running at a temperature below 15~mK. The Al sample boxes for both chips are mounted on the same cold-finger inside a single Cryoperm magnetic shield. A Radiall relay switch on the output lines of the two devices allows us to switch between measurements of one chip or the other. Supplementary Fig.~\ref{fig:fridgeSchematic} details the configuration of cabling, attenuation, filtering, and shielding inside the cryostat, as well as the room-temperature electronics hardware for control and readout. The inner surfaces of the Cryoperm magnetic shield and the mixing chamber shield were both coated with an infrared-absorbent layer \cite{sBarends2011}. 
For the charge biasing of the qubits, wiring limitations on our dilution refrigerator prevented us from connecting to all of the bias traces on the chips. For the non-Cu chip, charge bias lines are connected to $Q_B$ and $Q_C$; for the Cu chip, there is only a bias connection to $Q_A$.

\section{Supplementary Note 4: Device Parameters}

Supplementary Table~\ref{coincidence_table} lists relevant qubit parameters for both chips, including the qubit transition frequency $f_{01}$, the readout resonator frequency $f_{RO}$, the peak-to-peak maximum charge dispersion $2\delta f$, the mean and standard deviation from repeated baseline $T_1$ measurements, and the $E_{\rm J}/E_{\rm c}$ ratios. 
%
During the junction fabrication, the same double-angle evaporation process is used for the injector and qubit junctions, and thus all junctions on a chip have nominally the same critical current density. 
For each device, one of the junctions around the perimeter of the chip is connected to a 50-$\Omega$ bias lead to use as the injector junction (indicated by color highlighting in Supplementary Fig.~\ref{fig:chipImages}); the injector junction for each chip is $\sim$3 times the area of the qubit junctions. Because the junctions on the Cu and non-Cu chips were processed separately, the critical current densities on the two chips are slightly different. For the injector junctions, $R_n = 3.5~(3.0)~{\rm k}\Omega$ for the non-Cu (Cu) chips. The qubit junctions were all designed to have the same area and, based on witness junctions on the same chip written with the same area, had normal resistance of $R_n = 12.2~(10.8)~{\rm k}\Omega$ for the non-Cu (Cu) chips.

\begin{table*}[!htbp]
\begin{tabular}{ |p{1.5cm}||p{1.5cm}|p{1.7cm}|p{1.5cm}|p{1.6cm}|p{1.6cm}|p{1.6cm}|p{1.6cm}|}
 \hline
 \multicolumn{8}{|c|}{Qubit Parameters} \\
 \hline
 \hfil Device & \hfil Qubit & \hfil $f_{01} {\rm (GHz)}$ & \hfil $f_{RO} {\rm (GHz)}$ & \hfil $T_1 (\mu{\rm s)}$ & \hfil $\sigma(T_1) (\mu{\rm s)}$ & \hfil $2\delta f{\rm(MHz)}$  & \hfil $E_{\rm J}/E_{\rm c}$\\
 \hline

 \hfil\multirow{3}*{non-Cu}&  \hfil $Q_A$ & \hfil 4.6555   & \hfil 6.0431 & \hfil 34 & \hfil 10 & \hfil 3.743 & \hfil 24\\
                          &  \hfil $Q_B$ & \hfil 4.7363 & \hfil 6.1506 & \hfil 20 & \hfil 2 & \hfil 3.201 & \hfil 26\\
                          &  \hfil $Q_C$ & \hfil 4.8408   & \hfil 6.229  & \hfil 16 & \hfil 2 & \hfil 4.631 & \hfil 25\\
 \hline
 
 \hfil\multirow{3}*{Cu}&  \hfil $Q_A$ & \hfil 4.9959 & \hfil 6.3977 & \hfil 16 & \hfil 3 & \hfil 1.878 & \hfil 29\\
                          &  \hfil $Q_B$ & \hfil 5.2536 & \hfil 6.4868 & \hfil 21 & \hfil 5 & \hfil 1.146 & \hfil 32\\
                          &  \hfil $Q_C$ & \hfil 5.3190 & \hfil 6.5963 & \hfil 13 & \hfil 4 & \hfil 1.938 & \hfil 31\\

\hline
\end{tabular}
\caption{{\bf Qubit parameters for both non-Cu and Cu samples.}}
\label{coincidence_table}
\end{table*}




\section{Supplementary Note 5: Details of $\Delta\Gamma_1$ measurements}

For measurements of enhancements to the qubit relaxation rate following pulsing of the injector junction, in the main paper we present measurements of $\Delta\Gamma_1$ for $Q_C$ on the non-Cu chip and $Q_B$ for the Cu chip. In this section, we compile these measurements for the other qubits, and show that the response of the other qubits on each chip is consistent with the representative measurements in the main paper. 

\begin{figure}[!htbp]
\centering
\includegraphics[width=7in]{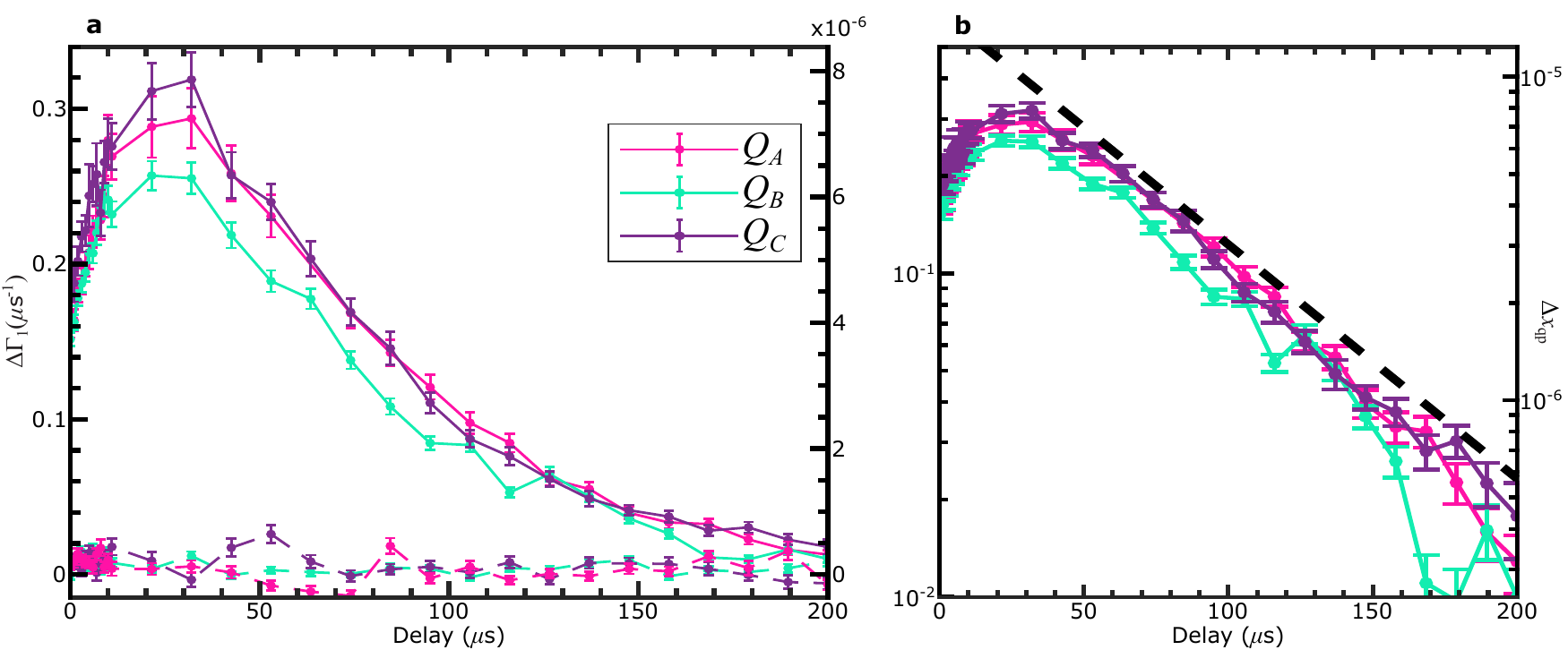}
  \caption{{\bf Enhancement of relaxation rate with controlled poisoning.} (a) Plot of $\Delta\Gamma_1$ vs. delay between injection pulse and $X$ pulse for $T_1$ measurement for all three qubits on both chips at $V_b$ = 1~mV where the Cu qubits are indicated by dashed lines joining the data points, while solid lines indicate the non-Cu qubits; the color labeling for each qubit is shown in the legend. (b) Plot of the same non-Cu data as in (a) plotted on a semilog scale; black dashed line indicates a characteristic decay time constant of 60~$\mu$s. The error bars on $\Delta\Gamma_1$ values represent 95\% confidence intervals from $T_1$ fits.
  \label{fig:Deltagamma_delay}}
\end{figure}

Supplementary Fig.~\ref{fig:Deltagamma_delay}(a) contains measurements of $\Delta\Gamma_1$ vs. the delay between the 10-$\mu$s injection pulse and the $X$ pulse for the relaxation measurement for all three qubits on both the non-Cu and Cu chips. In Supplementary Fig.~\ref{fig:Deltagamma_delay}(b), we plot the same data for the three qubits on the non-Cu chip on a semilog scale. The black dashed line corresponds to a characteristic timescale of 60~$\mu$s for injected phonons to leave the chip following the phonon arrival peak. Error bars on $\Delta \Gamma_{1}$ values, here and in Fig.~2 in the main paper, are calculated from fit errors with 95$\%$ confidence intervals from $T_{1}$ fits with contributions added in quadrature.

\begin{figure}[!b]
\centering
\includegraphics[width=6.4in]{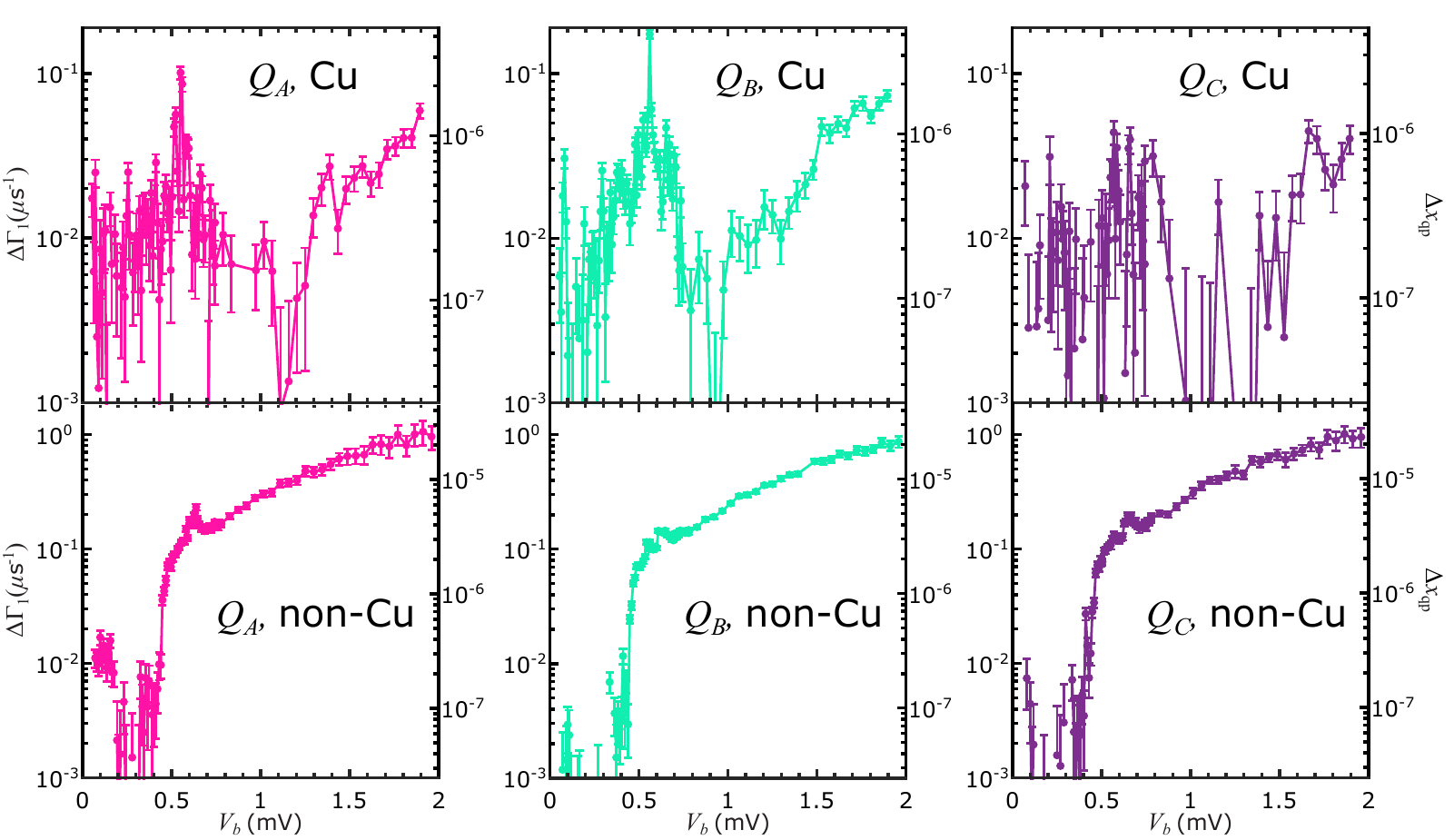}
  \caption{{\bf Enhancement of relaxation rate with controlled poisoning vs. injection amplitude.} Plots of $\Delta\Gamma_1$ vs. $V_b$ for a 30-$\mu$s delay for all qubits on the Cu sample (top row) and for the non-Cu sample (bottom row). The error bars on $\Delta\Gamma_1$ values represent 95\% confidence intervals from $T_1$ fits.
  \label{fig:Deltagamma_Vb1}}
\end{figure}

Supplementary Fig.~\ref{fig:Deltagamma_Vb1} contains plots of $\Delta\Gamma_1$ vs. $V_b$ for a delay of 30~$\mu$s for all three qubits on both chips. In Supplementary Fig.~\ref{fig:Deltagamma_Vb2}, we plot the same type of measurements but with a delay of 100~ns. In this second case, the antenna-resonance peaks from the photonic coupling to the Josephson radiation emitted by the injector junction are enhanced, while the remaining phononic poisoning is somewhat lower, as not all of the injected phonons have reached the qubit yet. The change in reduced QP density in the qubit junction leads, $\Delta x_{\rm qp}$, that is plotted on the right axes can be calculated from $\Delta\Gamma_1$ as $\Delta x_{\rm qp}=\pi\Delta\Gamma_1/\sqrt{2\Delta_{\rm Al}\omega_{01}/\hbar}$ \cite{sWang2014}, where $\omega_{01}$ is the qubit transition frequency. 

When the injection pulse amplitude is below $2\Delta_{\rm Al}$, we observe only minimal reduction in $T_1$; there is still some non-zero, but small, poisoning in this regime because our junction biasing scheme still permits the injector junction to undergo relaxation oscillations for small bias voltages \cite{sVernon1968}, 
where the junction can momentarily switch out to the gap before retrapping.

\vspace{20mm}

\newpage

\begin{figure}[!htbp]
\centering
\includegraphics[width=6.4in]{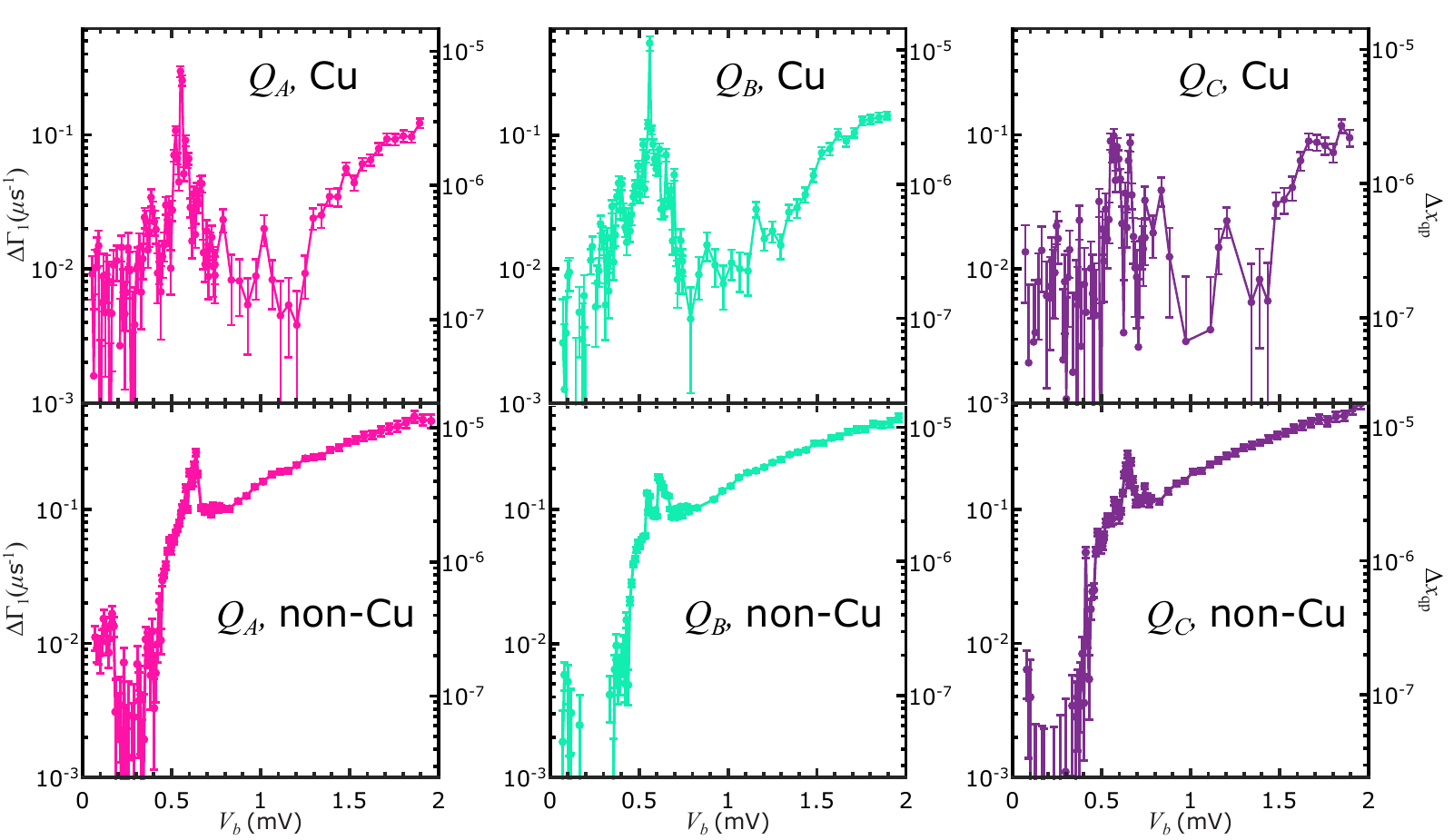}
  \caption{{\bf Enhancement of relaxation rate with controlled poisoning vs. injection amplitude.} 
  Plots of $\Delta\Gamma_1$ vs. $V_b$ with a 100-ns delay for all qubits on the Cu sample (top row) and for the non-Cu sample (bottom row). The error bars on $\Delta\Gamma_1$ values represent 95\% confidence intervals from $T_1$ fits.
  \label{fig:Deltagamma_Vb2}}
\end{figure}


\begin{figure}[!h]
\centering
\includegraphics[width=6.5in]{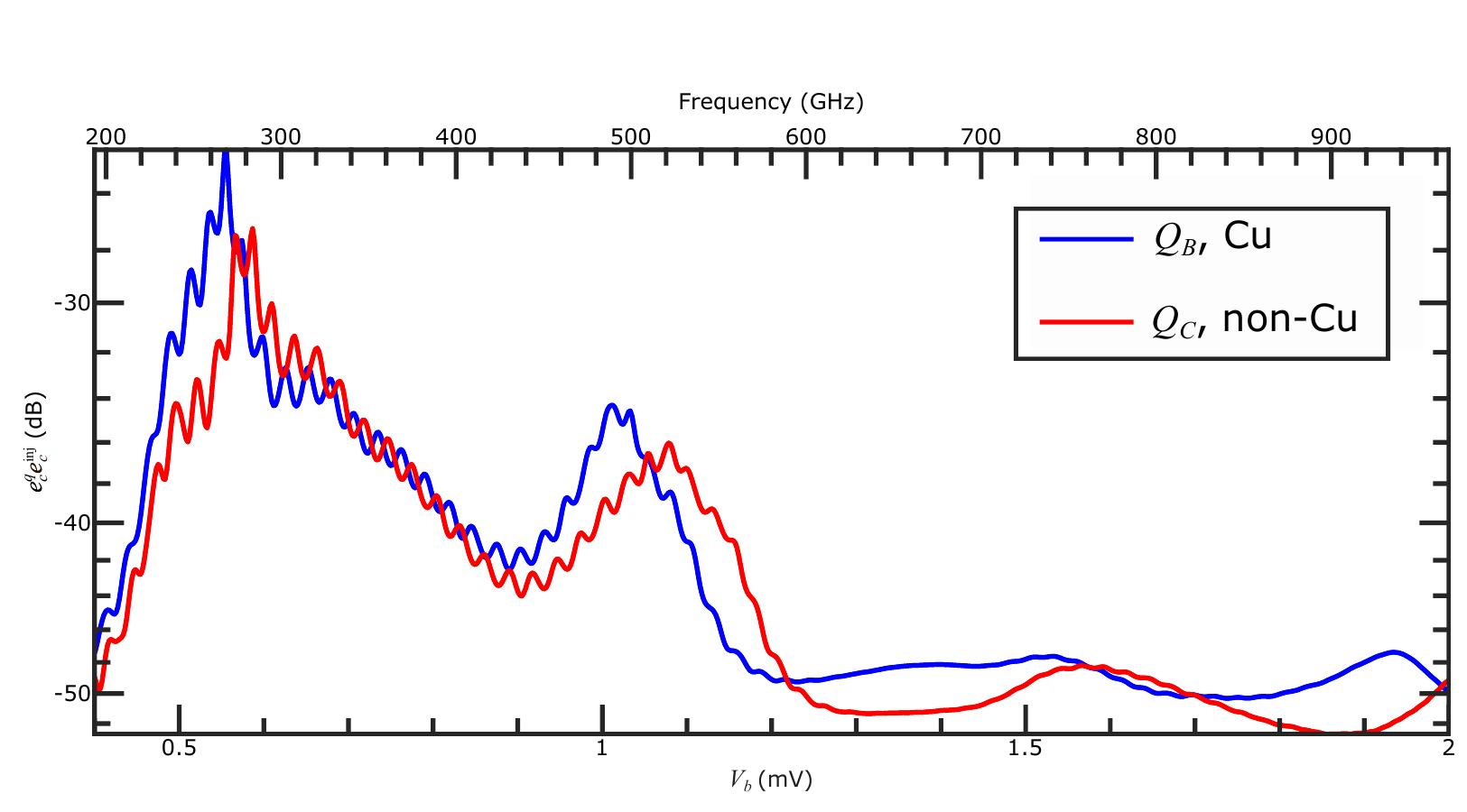}
  \caption{{\bf Antenna mode simulations.} Plot of $e_c^q e_c^{\rm inj}$ vs. $V_b$ for $Q_C$ on non-Cu chip and $Q_B$ on Cu chip, as described in the text; scale on upper axis converted to Josephson frequency of emitted radiation. 
  \label{fig:antenna}}
\end{figure}

\pagebreak[4]
\section{Supplementary Note 6: Antenna-mode simulations}

As described in the main paper, voltage bias of the injector junction will also induce poisoning from the emission of Josephson radiation. 
For a pulse amplitude $V_b$, the Josephson radiation will have frequency $V_b/\Phi_0$, where $\Phi_0 \equiv h/2e$ is the magnetic flux quantum; $h$ is Planck's constant and $e$ is the electron charge. 
Such electromagnetic radiation can be resonantly absorbed by qubit structures acting as antennas, with typical resonant frequencies in the hundreds of GHz range. The absorbed radiation can then drive high-frequency currents through the qubit junction and generate QPs, as described recently in Ref.~\cite{sRafferty2021,sliu2022}. A related photon-based QP poisoning mechanism was considered in Ref.~\cite{sHouzet2019}.

In order to model the spurious qubit antenna resonances on our devices, we follow the analysis in Ref.~\cite{sRafferty2021,sliu2022} and compute the radiation impedances of the injector junction and the qubit structure with a finite-element simulation using CST Microwave \cite{sCST}. 

With the critical current values for the injector and qubit junctions extracted from the on-chip witness junction measurements, we calculate the product of the coupling efficiencies to free space for the injector junction $e_c^{\rm inj}$ and the qubit junction $e_c^{q}$. In Supplementary Fig.~\ref{fig:antenna}, we plot this product as a function of the injector junction pulse amplitude $V_b$ for both $Q_B$ on the Cu chip and $Q_C$ on the non-Cu chip. The fundamental peaks in the simulation for both qubits match the measured antenna resonances from $\Delta\Gamma_1$ in Fig.~2(b) in the main paper and Supplementary Fig.~\ref{fig:Deltagamma_Vb2} in the supplement.

\section{Supplementary Note 7: Parity switching power spectra for all qubits} \label{PSD-sec}

\begin{figure}[!b]
\centering
\includegraphics[width=6.8in]{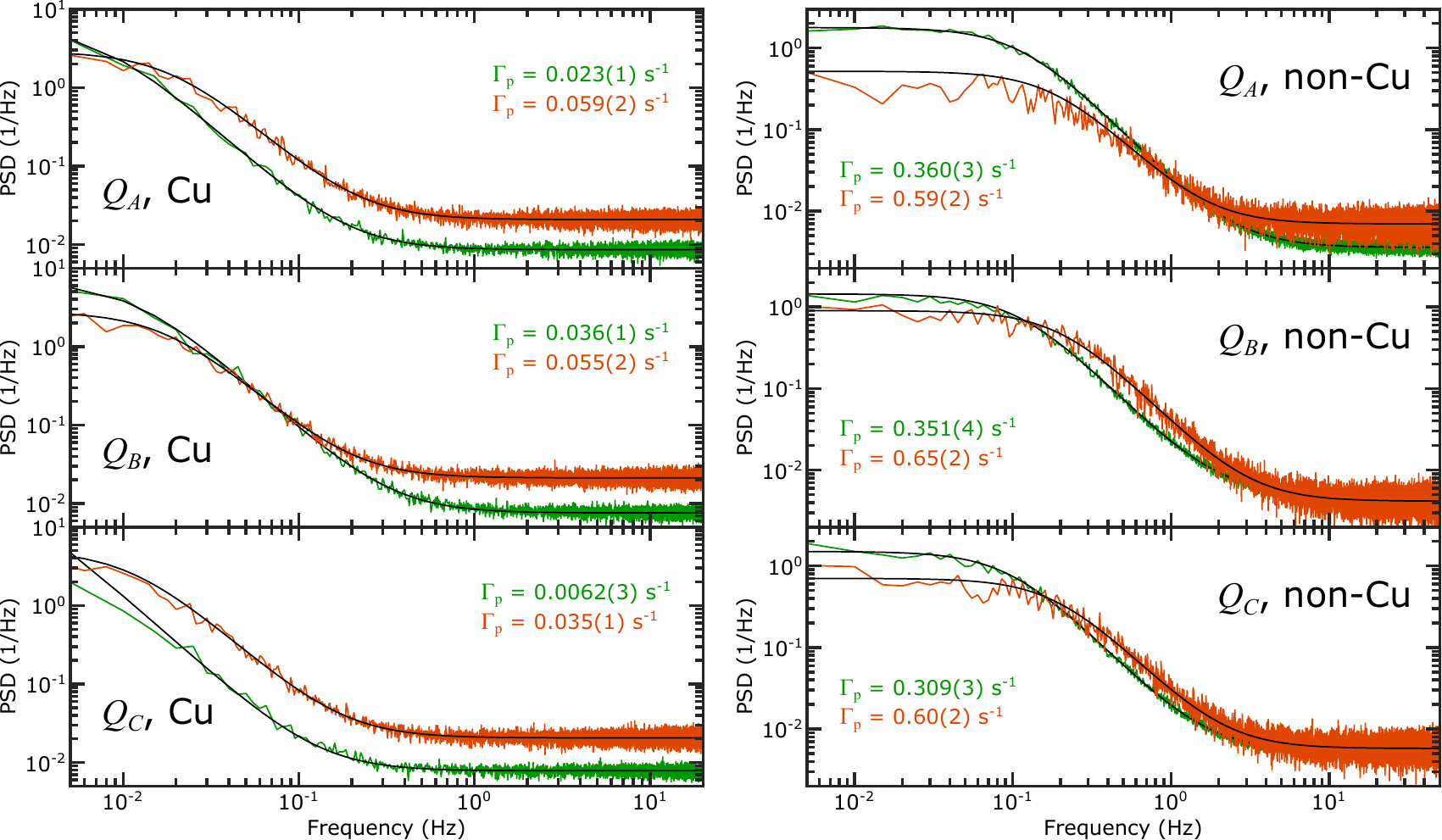}
  \caption{{\bf QP parity switching power spectra.} Plots of power spectral densities of QP parity switching for all qubits measured on the first (green) and second (orange) cooldowns; black lines correspond to PSD fits using Supplementary Eq. (\ref{lorentzian}). The error bars on $\Gamma_{\rm p}$ values represent 95\% confidence intervals from the fits. 
  \label{fig:PSD-all}}
\end{figure}

We implement a Ramsey pulse sequence that has been used previously to map QP parity onto qubit 1-state occupation \cite{sRiste2013, sSerniak2018, sChristensen2019}). 
We apply an $X/2$ pulse, idle for a time corresponding to a quarter of a qubit precession period, then apply a $Y/2$ pulse, followed by a qubit measurement. 
If the offset charge corresponds to the point of maximum charge dispersion, the final $Y/2$ pulse will rotate the state vector to the north/south poles of the Bloch sphere dependent on the QP parity state. 
Although some of the qubits on each chip have connections to the charge-bias line, we have chosen to perform our QP parity switching measurements without active stabilization of the offset charge. This allows the QP parity measurements to proceed without interruptions from periodic charge-tomography sequences \cite{sChristensen2019}. However, 
when the offset charge jumps to near $(n+1/2)e$ (for integer $n$), where the bands cross, the fidelity of the QP parity-mapping sequence approaches zero.

In order to compute the power spectral density of the QP parity switching, we perform the QP parity switching measurement on each qubit with 20,000 single shots at a repetition period of 10~ms (although PSD measurements for the Cu chip on the second cooldown used a 25-ms repetition period). For each single-shot measurement stream, we apply a simple thresholding based on the 0/1 readout calibration levels for each respective qubit to produce a digital time trace of QP parity. We then compute the PSD from the resulting digital trace and average several such PSD traces together (between 20-160) to obtain the curves in Fig.~3(a) in the main paper and Supplementary Fig.~\ref{fig:PSD-all}. Because we are not actively stabilizing the offset charge at the point of maximum dispersion, some of the PSD traces that are being averaged will have the environmental offset charge near the degeneracy point, where the QP parity readout fidelity vanishes. This results in an enhancement of the white noise floor, but still allows for a clear extraction of the characteristic QP parity switching rate.

We are able to fit the resulting power spectra with a single Lorentzian using the form described in Ref.~\cite{sRiste2013}:
\begin{equation}
    S_p(f)=\frac{4F^2\Gamma_{\rm p}}{(2\Gamma_{\rm p})^2+(2\pi f)^2}+(1-F^2)\Delta t,
\label{lorentzian}
\end{equation}
where $\Gamma_{\rm p}$ is the parity switching rate, $F$ is the parity sequence mapping fidelity, and $\Delta t$ is the parity measurement repetition period.

\renewcommand{\sectionautorefname}{12}

Supplementary Fig.~\ref{fig:PSD-all} shows the PSD for all three qubits on each chip. During our experiment, after collecting a majority of our data once the dilution refrigerator had been cold for several months, an unplanned power outage caused our dilution refrigerator to warm up to room temperature. Upon immediately cooling the same two devices back down, without making any changes to the wiring or shielding, we remeasured the PSD for each qubit within a few weeks of the start of this second cooldown. The plots in Supplementary Fig.~\ref{fig:PSD-all} contain the PSD for each qubit on both chips measured on the first and second cooldowns. For all qubits, the QP parity switching rates increase on the second cooldown, likely because some elements in the qubit environment, for example, the isolators, attenuators, or shields, have not yet fully cooled to the base temperature (see Supplementary Note \autoref{sec:cooldowns}\hspace*{-1mm} for further discussion). Nonetheless, the $\Gamma_{\rm p}$ values on the Cu chip remain at least one order of magnitude lower compared to the non-Cu chip.

\section{Supplementary Note 8: QP Parity switching with pulsed injection for all qubits}

\begin{figure}[!h]
\centering
 \includegraphics[width=6.8in]{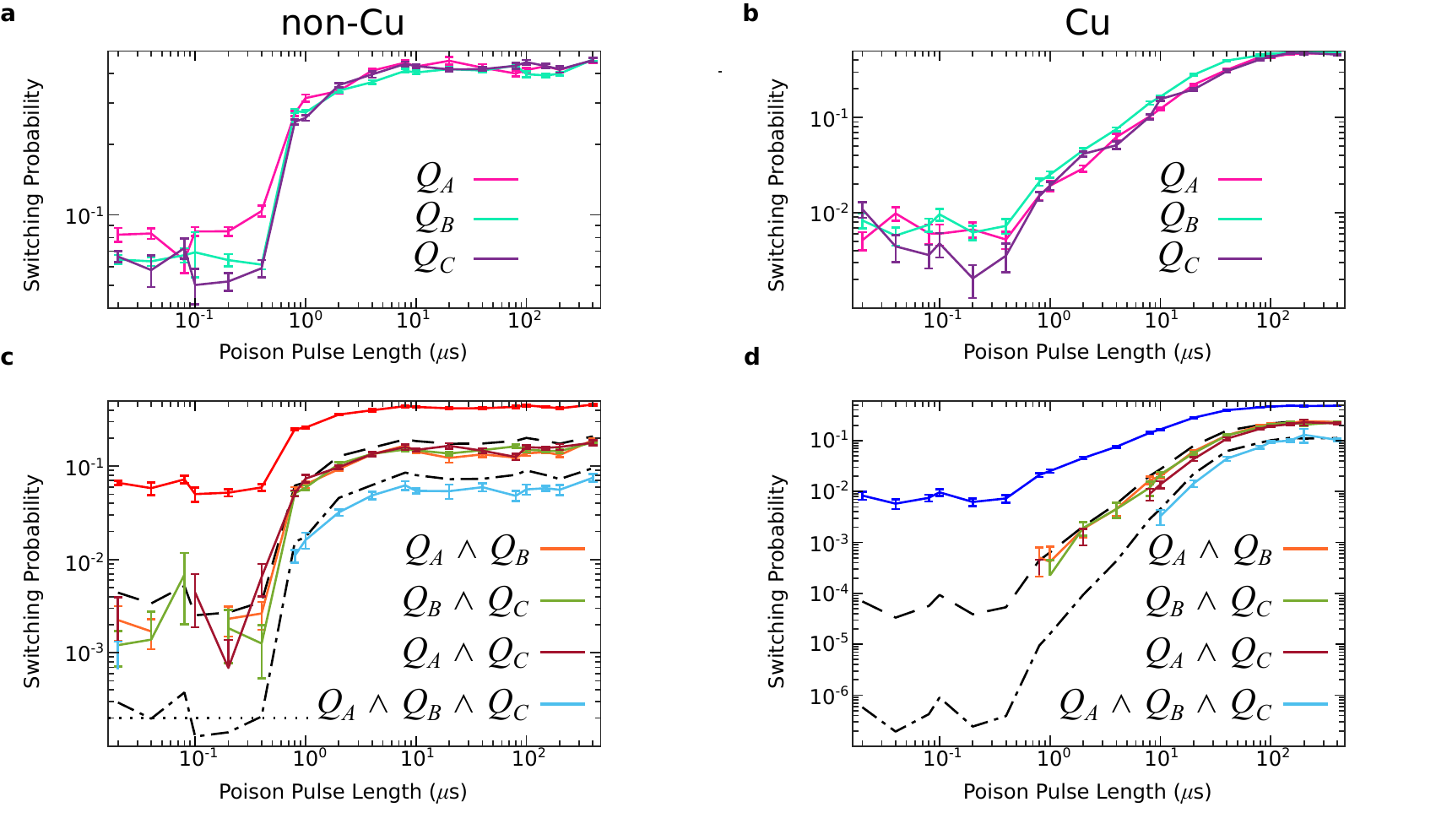}
  \caption{{\bf QP parity measurements with controlled injection for all qubits.} The switching probability, given by the measured parity switching rate divided by the pulse rate of the injector junction (20~Hz) vs. poison pulse duration for all three qubits on the (a) non-Cu and (b) Cu chips. Injection pulse amplitude $V_b$ is 1~mV for both sets of measurements. The probability of double and triple coincidence events for the (c) non-Cu and (d) Cu samples are shown in comparison to the single switching probabilities [$Q_C$ on non-Cu chip (red), $Q_B$ on Cu chip (blue)] and the square (dash) and cube (dot-dash) of the single switching probabilities. The expected random background probability for two-fold coincidences on the non-Cu chip is indicated by the dotted line in (c). Error bars computed from standard Poisson counting errors.
  \label{fig:PSwitch}}
\end{figure}

To complement Fig.~3(b) in the main paper, in this section, we plot the measured switching probability for each qubit on the non-Cu chip [Supplementary Fig.~\ref{fig:PSwitch}(a)] and Cu chip [Supplementary Fig.~\ref{fig:PSwitch}(b)] as a function of the injector junction pulse length. For each chip, all three qubits exhibit a similar behavior. Supplementary Fig.~\ref{fig:PSwitch}(c,d) contain the double- and triple-coincidence switching probabilities for the non-Cu and Cu chips, along with comparisons to the square and cube of the single-qubit switching probabilities for one of the qubits on each chip, as discussed in the main paper.

\section{Supplementary Note 9: Identification of QP parity switching events}

\begin{figure}[h!]
\centering
\includegraphics[width=7in]{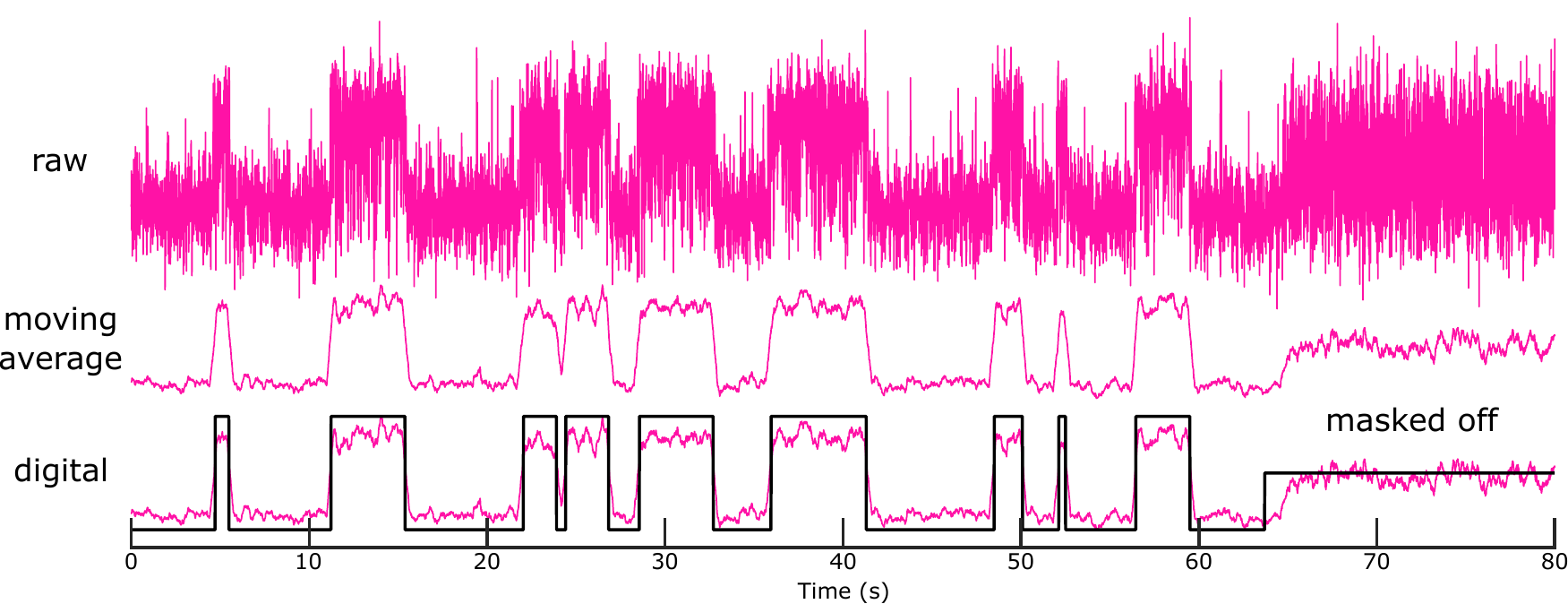}
  \caption{{\bf Identification of QP parity switching events.} Starting with the raw data at the top, a moving average over 40 time steps is then applied (middle); this is then converted to a digital signal using an HMM approach, shown as the black line on top of the averaged data in the trace at the bottom. For the final $\sim$17~s, the environmental offset charge for the qubit was near charge degeneracy. The analysis code accounts for this by masking off this portion of the data, which is reflected in the digital signal displayed halfway between the levels for the different parities. This data is taken from $Q_A$ for the non-Cu chip.}
  
  
  \label{fig:parity-details}
\end{figure}

In order to locate the parity switching steps from the simultaneous QP parity measurements, we apply the following data processing steps. First, because the offset charge was not actively stabilized, we need to identify the portions of the data stream for each qubit where the environmental offset charge jumped to near the degeneracy point, where the parity mapping fidelity approaches zero.
This involves finding the envelope of the peak-to-peak signal for the parity time trace with an applied moving average of 100 time steps. If the envelope is below a threshold determined by the qubit 0/1 readout calibration levels, the portion of data until the envelope extends above the threshold is masked off and not analyzed further when digitizing the parity time traces. We next digitize the parity time traces by applying a moving average to the unmasked raw parity data to improve the signal-to-noise ratio. We then use a hidden Markov model (HMM) to identify the parity states. For the QP parity data without junction injection presented here, we use a moving average of 40 time steps. After fitting Gaussians to the qubit 0/1 single-shot readout calibration measurements, 
we use these distributions to assign a probability for the parity signal to have a value along the signal axis corresponding to an odd- or even-parity state. For the HMM, we also set the probability for the system to transition from odd to even parity and vice versa based on the repetition time of the single shots and the $\Gamma_{\rm p}$ extracted separately from the QP parity PSDs for each qubit.  With 
this information, we then use the Viterbi algorithm to fit a digital signal to the averaged data, thus extracting the most probable parity state given the readout value along the signal axis. In a few instances, we use a modified HMM scheme for the parity analysis. This involves implementing a simple threshold method which assigns the parity of the state based on the data with an applied moving average relative to the total mean of the data. With a parity value assigned at each time index, we derive the statistics for the value of the parity signal given its state. We then use these parameters and the transition probabilities described previously to augment the HMM approach and fit a digital signal to the averaged data.
Supplementary Fig.~\ref{fig:parity-details} shows an example of ths parity switching analysis for $Q_A$ on the non-Cu chip. 

We then use the digital signal that was found through the HMM scheme to locate parity switches. We take the absolute value of the difference of adjacent points of the digital signal, which results in a peak at the location of each parity switch. The parity switching rate for each qubit $r_i$, where $i=A,B,C$, is given by $N_i/\tau_i$, where $N_i$ is the total number of parity switches for that particular qubit and $\tau_i$ is the total duration of unmasked data for the qubit. The uncertainty in $r_i$ comes from the standard Poisson counting errors $N_i^{1/2}/\tau_i$.

\section{Supplementary Note 10: Extraction of QP parity switching coincidences}

Measuring the parity of all three qubits on either chip simultaneously allows us to track correlated events between qubits. In order to identify parity switching coincidences, we must process the digital parity traces obtained as described in the previous section to look for simultaneous switching between qubits. Because the moving averages that are applied to the raw parity measurement data to improve the signal-to-noise ratio also cause the switching events to have a shallower step, we must implement a windowing process to find coincidences. Because the effective width of the switching steps is approximately equal to the number of moving average time steps, we set our window size to match the number of moving averages, 
thus, coincident switches should occur no farther apart than the width of the falling/rising edges. 

\begin{figure}[!htbp]
\centering
\includegraphics[width=6.8in]{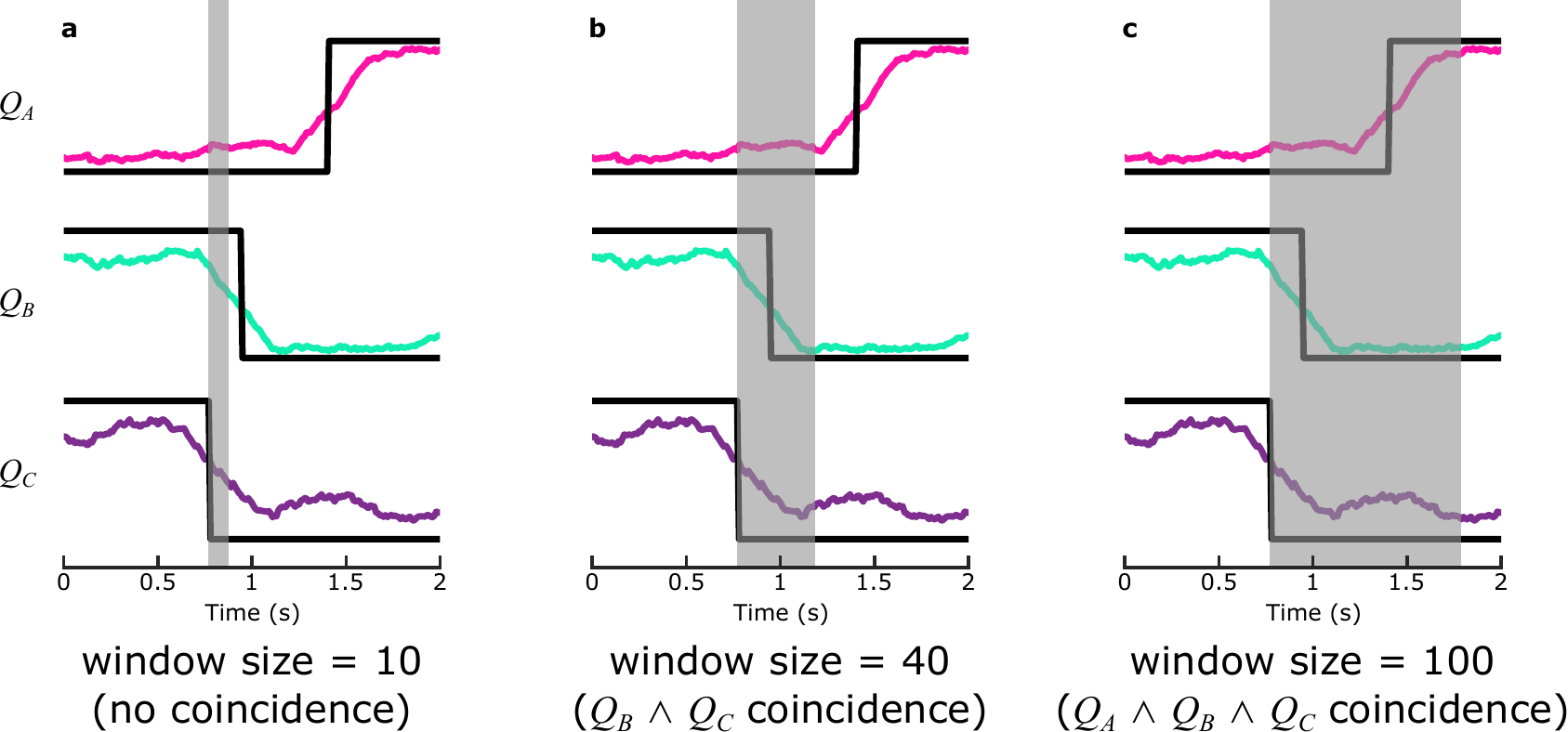}
  \caption{{\bf Data windowing for coincidence identification.} Example section of simultaneous QP parity data for non-Cu chip for a moving average over 40 time steps, with different window sizes applied. Window size indicated by shaded gray rectangle of (a) 10 time steps, resulting in no identified coincidences; (b) 40 time steps, correctly identifying a $Q_B \land Q_C$ coincidence; (c) 100 time steps, misidentifying a $Q_A\land Q_B\land Q_C$ coincidence. 
  \label{fig:windowSize}}
\end{figure}

Supplementary Fig.~\ref{fig:windowSize} shows the effects of different window sizes for the same example parity data trace. We sweep our window through the simultaneous digital signals, and if multiple switches occur within our window size, they are identified as coincidences.
For a window size well below the number of moving averages, the code misses a double coincidence [Supplementary Fig.~\ref{fig:windowSize}(a)], while for a window size much greater than the number of moving averages, switches from separate events are misidentified as a coincidence. 

Following the coincidence switching identification, the events are indexed with the appropriate type ($Q_A\land Q_B$, $Q_B\land Q_C$, $Q_A\land Q_C$, or $Q_A\land Q_B\land Q_C$). With this approach, every triple coincidence is also counted as three double coincidences. We also restrict each switch of a given qubit to participate in only one event per coincidence type. For example, a $Q_B$ switch cannot be used for two $Q_A\land Q_B$ coincidences, but could be used for a $Q_A\land Q_B$ coincidence and a $Q_B\land Q_C$ coincidence. 
In Supplementary Fig.~\ref{fig:coincidence-details}, we present example simultaneous parity traces for all three qubits for both chips. We also represent the locations of extracted coincidences with vertical dashed lines.

The switching rate for each type of coincidence event $r_i$, where $i=AB,BC,AC,ABC$, is given by $N_i/\tau_i$, where $N_i$ is the total number of events and $\tau_i$ is the total duration of unmasked data for event type $i$. Note that double coincidences between qubits $j$ and $k$ are only counted during the period when both qubits are unmasked; similarly, triple coincidences require that all three qubits are unmasked. The uncertainty in $r_i$ comes from the standard Poisson counting errors $N_i^{1/2}/\tau_i$.

In Supplementary Table~\ref{observed_rates_table_windowSize}, we explore the effect of different window sizes and moving averages on the observed parity switching rates for both chips. For higher averages, we observe a moderate decrease ($\sim$10\%) in the single-qubit switching rate, which we attribute to occasional narrow features with two closely spaced switches that get averaged below the threshold for larger numbers of moving averages. At the same time, the double- and triple-coincidence rates increase somewhat as the window size increases. Nonetheless, we still observe the same overall trend between the two chips: the single-qubit parity switching rates for the Cu chip remain $\sim$1 order of magnitude lower than for the non-Cu chip, and the double- and triple-coincidence rates are still $\sim$2 orders of magnitude lower.

\begin{figure}[!htbp]
\centering
\includegraphics[width=6.8in]{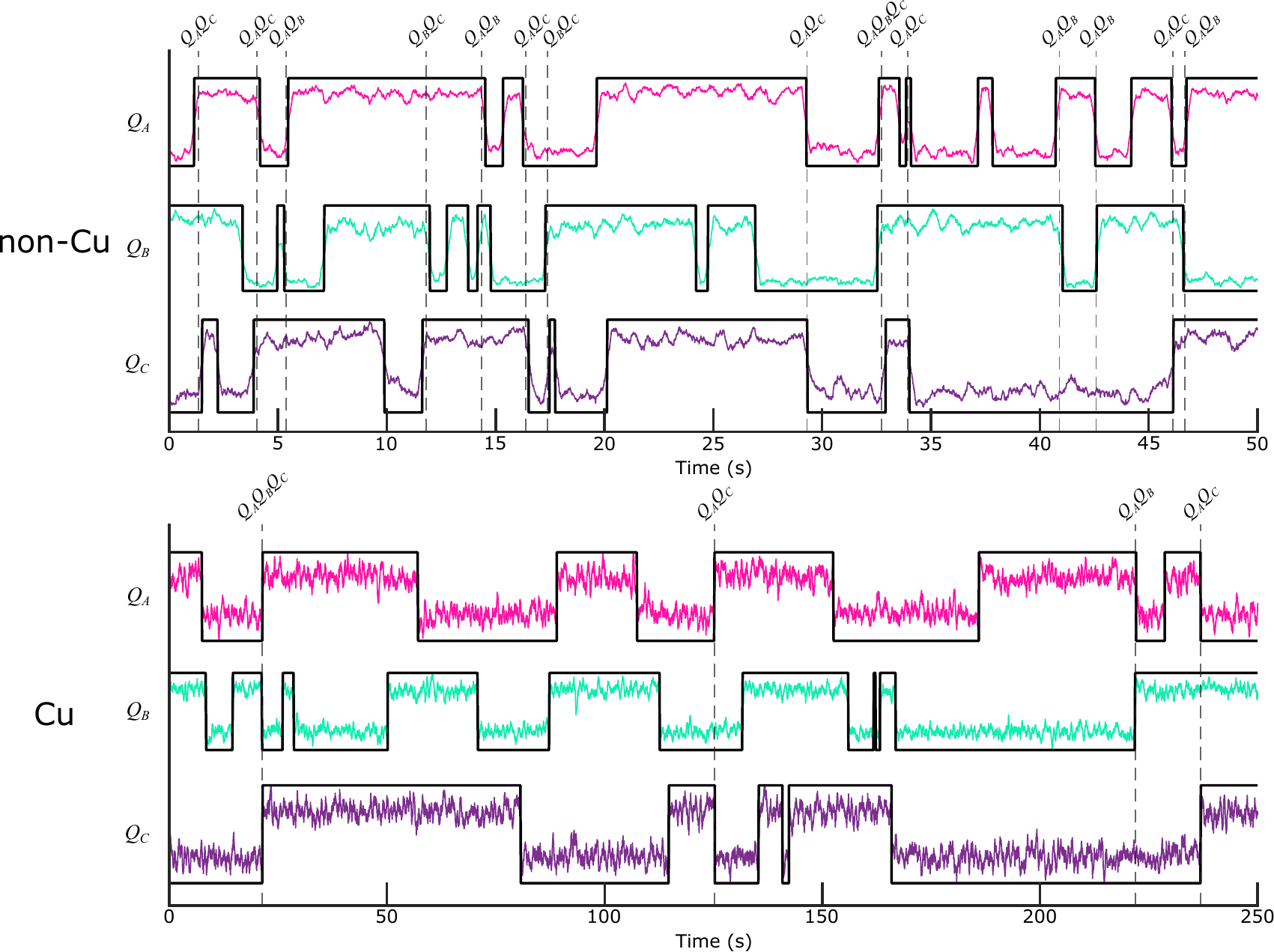}
  \caption{{\bf Identification of QP parity switching coincidences for both chips.} Examples of extracted digital QP parity signals from simultaneous QP parity data and identification of coincidences. Note the 5x difference in the timespans between the bottom plot for the Cu chip and the upper plot for the non-Cu chip.
  \label{fig:coincidence-details}}
\end{figure}


\begin{table*}
\begin{tabular}{|p{1.1cm}||p{2.2cm}|p{2.1cm}|p{1.32cm}|p{1.32cm}|p{1.32cm}|p{1.5cm}|p{1.5cm}|p{1.5cm}|p{2.2cm}|}
 \hline
 \multicolumn{10}{|c|}{\makecell[b]{Observed, random background, and extracted parity switching rates ($s^{-1}\times 10^{-3}$) \\ for different window sizes and moving averages}}\\
 \hline
 \hfil Device & \hfil \makecell{window size, \\ moving average} & \hfil Type & \hfil $Q_A$ & \hfil $Q_B$ & \hfil $Q_C$ & \hfil $Q_A\land Q_B$ & \hfil $Q_B\land Q_C$ & \hfil $Q_A\land Q_C$ & \hfil $Q_A\land Q_B\land Q_C$\\
 \hline
 \hfil\multirow{9}*{non-Cu} & \hfil\multirow{3}*{20, 20} & \hfil observed & \hfil 320(3) & \hfil 333(4) & \hfil 272(3) & \hfil 42(2) & \hfil 38(2) & \hfil 36(1) & \hfil 6.1(8)\\\cline{3-10}
                            & \hfil & \hfil background & \hfil - & \hfil - & \hfil - & \hfil 21.3(4) & \hfil 18.1(3) & \hfil 17.4(3) & \hfil 1.16(2)\\\cline{3-10}
                            & \hfil & \hfil extracted & \hfil 410(20) & \hfil 430(20) & \hfil 320(20) & \hfil 110(10) & \hfil 100(10) & \hfil 100(10) & \hfil 23(8)\\\cline{2-10}
                & \hfil\multirow{3}*{20, 40} & \hfil observed &     \hfil 299(3) & \hfil 301(4) & \hfil 252(3) & \hfil 37(2) & \hfil 35(2) & \hfil 33(1) & \hfil 5.2(8)\\\cline{3-10}
                            & \hfil & \hfil background & \hfil - & \hfil - & \hfil - & \hfil 18.0(3) & \hfil 15.2(3) & \hfil 15.0(2) & \hfil 0.91(2)\\\cline{3-10}
                            & \hfil & \hfil extracted & \hfil 390(20) & \hfil 390(20) & \hfil 300(20) & \hfil 100(10) & \hfil 100(10) & \hfil 90(10) & \hfil 20(7)\\\cline{2-10}
                & \hfil\multirow{3}*{40, 40} & \hfil observed & \hfil 299(3) & \hfil 301(4) & \hfil 252(3) & \hfil 65(2) & \hfil 60(2) & \hfil 57(2) & \hfil  12(1)\\\cline{3-10}
                            & \hfil & \hfil background & \hfil - & \hfil - & \hfil - & \hfil 36.0(6) & \hfil 30.3(5) & \hfil 30.1(5) & \hfil 3.62(8)\\\cline{3-10}
                            & \hfil & \hfil extracted & \hfil 200(20) & \hfil 190(20) & \hfil 120(20) & \hfil 180(10) & \hfil 170(10) & \hfil 150(10) & \hfil 64(9)\\\cline{2-10}
 \hline \hline
 \hfil\multirow{9}*{Cu} & \hfil\multirow{3}*{20, 20} & \hfil observed & \hfil 25.9(4) & \hfil 36.5(5) & \hfil 32.7(7) & \hfil 0.56(8) & \hfil 0.6(2) & \hfil 0.5(1) & \hfil 0.06(6)\\\cline{3-10}
                            & \hfil & \hfil background & \hfil - & \hfil - & \hfil - & \hfil 0.189(4) & \hfil 0.239(6) & \hfil 0.169(4) & \hfil 0.00124(4)\\\cline{3-10}
                            & \hfil & \hfil extracted & \hfil 49(1) & \hfil 70(1) & \hfil 63(2) & \hfil 1.1(7) & \hfil 1.2(9) & \hfil 0.7(7) & \hfil 0.5(5)\\\cline{2-10}
                & \hfil\multirow{3}*{20, 40} & \hfil observed & \hfil 22.1(3) & \hfil 33.6(5) & \hfil 23.0(5) & \hfil 0.57(8) & \hfil 0.4(1) & \hfil 0.36(9) & \hfil 0.06(6)\\\cline{3-10}
                            & \hfil & \hfil background & \hfil - & \hfil - & \hfil - & \hfil 0.149(3) & \hfil 0.155(4) & \hfil 0.102(3) & \hfil 0.00069(2)\\\cline{3-10}
                            & \hfil & \hfil extracted & \hfil 42(1) & \hfil 65(1) & \hfil 44(1) & \hfil 1.2(7) & \hfil 0.7(8) & \hfil 0.6(7) & \hfil 0.5(5)\\\cline{2-10}
                & \hfil\multirow{3}*{40, 40} & \hfil observed & \hfil 22.1(3) & \hfil 33.6(5) & \hfil 23.0(5) & \hfil 0.8(1) & \hfil 0.8(2) & \hfil 0.5(1) & \hfil 0.06(6)\\\cline{3-10}
                            &  \hfil & \hfil background & \hfil - & \hfil - & \hfil - & \hfil 0.298(6) & \hfil 0.310(9) & \hfil 0.204(6) & \hfil 0.00274(9)\\\cline{3-10}
                            & \hfil & \hfil extracted & \hfil 41(1) & \hfil 63(1) & \hfil 43(2) & \hfil 1.9(7) & \hfil 1.6(9) & \hfil 1.0(7) & \hfil 0.4(6)\\\cline{2-10}
 \hline
\end{tabular}
\caption{{\bf Comparison of rates for different windowing and averaging.} Observed switching rates, random background coincidence rates, and extracted poisoning event rates for different window size and moving average combinations across both chips. The entries for window size = 40 and moving average = 40 match those in Fig. 4(b) of the main paper. Note the scale factor of $10^{-3}$ on the rate units.}
\label{observed_rates_table_windowSize}
\end{table*}

\section{Supplementary Note 11: Identification of correlated QP poisoning rates}

For a set of observed single-qubit parity switching rates $r_{A-C}^{\rm obs}$ with a non-zero window $\Delta t$ for identifying double- and triple-coincidence switching, one would expect a rate for random uncorrelated coincidence switching given by the product of the probabilities for observing a parity switch for each of the constituent qubits in the coincidence event during the interval $\Delta t$. Thus, the expected random background double-concidence rate for qubits $i$ and $j$ is given by $(r_i^{\rm obs}\Delta t) (r_j^{\rm obs}\Delta t)/\Delta t$; similarly the expected random background triple-coincidence rate for qubits $i$, $j$, and $k$ is given by $(r_i^{\rm obs}\Delta t) (r_j^{\rm obs}\Delta t) (r_k^{\rm obs}\Delta t)/\Delta t$. These expected random coincidence parity switching rates are listed in Supplementary Table~\ref{observed_rates_table_windowSize} for different numbers of moving averages and window sizes. The error bars for these random background coincidence rates were computed by summing the fractional uncertainty for each observed rate in quadrature. We note that these random background coincidence rates remain well below the observed rates as we vary the windowing and averaging.


While the quantities we measure in our simultaneous parity measurements are the observed parity switching rates, we would like to compute the actual poisoning event rates for each qubit, or group of qubits in the case of correlated poisoning. This calculation requires accounting for the random background coincidence switching described above, as well as the probability for recording a parity switch for a given poisoning event: 1/2 in the case of single-qubit poisoning, 1/4 for double-qubit correlated poisoning, and 1/8 for triple-qubit poisoning, as described in the main paper.

For the observed parity switching, each double-coincidence event will also be recorded as two single-qubit switching events; similarly, each triple-coincidence event will also be recorded as three double-qubit switching events and three single-qubit switching events. Here, we define the extracted poisoning event rates $r_i$ to be exclusive; for example, a 
single poisoning event that couples to both $Q_A$ and $Q_B$ will contribute to $r_{AB}$ but will not contribute to $r_A$ or $r_B$. 

Based on these criteria, we can use the observed parity switching rates $r_i^{\rm obs}$ to compute the probability for observing each type of parity switching event in a window interval $\Delta t$ as $p_i^{\rm obs}=r_i^{\rm obs}\Delta t$. We can then derive expressions for the probability of observing each type of parity switching event in terms of the actual probability for each type of poisoning event:
\begin{align}
\begin{aligned}
p_A^{\rm obs} & = \frac{1}{2}\left(p_{ABC}+p_{AB}+p_{AC}+p_A \right) \\
p_B^{\rm obs} & = \frac{1}{2}\left(p_{ABC}+p_{AB}+p_{BC}+p_B \right) \\
p_C^{\rm obs} & = \frac{1}{2}\left(p_{ABC}+p_{AC}+p_{BC}+p_C \right) \\
p_{AB}^{\rm obs} & = \frac{1}{4}\left(p_{ABC}+p_{AB}+p_Ap_B \right) \\
p_{BC}^{\rm obs} & = \frac{1}{4}\left(p_{ABC}+p_{BC}+p_Bp_C \right) \\
p_{AC}^{\rm obs} & = \frac{1}{4}\left(p_{ABC}+p_{AC}+p_Ap_C \right) \\
p_{ABC}^{\rm obs} & = \frac{1}{8}\left(p_{ABC}+p_Ap_Bp_C+p_{AB}p_C+p_Ap_{BC}+p_{AC}p_B \right). \\
\end{aligned}
\label{prob_eqs}
\end{align}

\noindent With the experimentally measured switching probabilities $p_i^{\rm obs}$, we numerically solve the system of equations to obtain the actual poisoning probabilities $p_i$. We then calculate the actual poisoning rates $r_i=p_i/\Delta t$. We compute the error bars on each actual poisoning probability by numerically computing the derivative with respect to each of the observed switching probabilities, then multiplying by the corresponding Poisson error bar for the observed switching probability and summing these together in quadrature.


In Supplementary Table~\ref{observed_rates_and_probs_table}, we list the values $N_i$ for each single qubit parity switch and coincidence event, as well as the total unmasked duration $\tau_i$ for the particular type of event. For the right three columns, the observed parity switching rates $r_i^{\rm obs}$, expected random background coincidence rates $r_i^{\rm background}$, and extracted actual poisoning rates $r_i$ correspond to the values presented in Fig.~4(b) in the main paper.


\begin{table*}
\begin{tabular}{ |p{1.1cm}||p{2.1cm}|p{2.1cm}|p{2.1cm}|p{2.1cm}|p{2.1cm}|p{2.6cm}|p{2.1cm}|}
 \hline
 \multicolumn{8}{|c|}{Observed parity switching and extracted poisoning event rates}\\
 \hline
 \hfil Device & \hfil Qubit(s) & \hfil Separation & \hfil $N_i$ & \hfil $\tau_i$ (s) & \hfil $r_{i}^{\rm obs}$ (s$^{-1}$) & \hfil $r_{i}^{\rm background}$ (s$^{-1}$) & \hfil $r_{i}$ (s$^{-1}$) \\
 \hline
 \hfil\multirow{7}*{non-Cu}& \hfil $Q_A$ & \hfil - & \hfil 8528 & \hfil 28,557 & \hfil 0.299(3) & \hfil - & \hfil 0.20(2)\\\cline{2-8}
                &  \hfil $Q_B$ & \hfil - & \hfil 5202 & \hfil 17,272 & \hfil 0.301(4) & \hfil - & \hfil 0.19(2)\\\cline{2-8}
                &  \hfil $Q_C$ & \hfil - & \hfil 7959 & \hfil 31,609 & \hfil 0.252(3) & \hfil - & \hfil 0.12(2)\\\cline{2-8}
                &  \hfil $Q_A\land Q_B$ & \hfil 5.3 mm & \hfil 832 & \hfil 12,851 & \hfil 0.065(2) & \hfil 0.0360(6) & \hfil 0.18(1)\\\cline{2-8}
                &  \hfil $Q_B\land Q_C$ & \hfil 4.5 mm & \hfil 670 & \hfil 11,124 & \hfil 0.060(2) & \hfil 0.0303(5) & \hfil 0.17(1)\\\cline{2-8}
                &  \hfil $Q_A\land Q_C$ & \hfil 2.0 mm & \hfil 1078 & \hfil 18,941 & \hfil 0.057(2) & \hfil 0.0301(5) & \hfil 0.15(1)\\\cline{2-8}
                &  \hfil $Q_A\land Q_B\land Q_C$ & \hfil - & \hfil 109 & \hfil 8,842 & \hfil 0.012(1) & \hfil 0.00362(8) & \hfil 0.064(9)\\\cline{2-8}
 \hline \hline
 \hfil\multirow{7}*{Cu}& \hfil $Q_A$ & \hfil - & \hfil 4031 & \hfil 182,103 & \hfil 0.0221(3) & \hfil - & \hfil 0.041(1)\\\cline{2-8}
                &  \hfil $Q_B$ & \hfil - & \hfil 4515 & \hfil 134,192 & \hfil 0.0336(5) & \hfil - & \hfil 0.063(1)\\\cline{2-8}
                &  \hfil $Q_C$ & \hfil - & \hfil 1779 & \hfil 77,322 & \hfil 0.0230(5) & \hfil - & \hfil 0.043(2)\\\cline{2-8}
                &  \hfil $Q_A\land Q_B$ & \hfil 5.3 mm & \hfil 66 & \hfil 78,936 & \hfil 0.0008(1) & \hfil 0.000298(6) & \hfil 0.0019(7)\\\cline{2-8}
                &  \hfil $Q_B\land Q_C$ & \hfil 4.5 mm & \hfil 20 & \hfil 25,376 & \hfil 0.0008(2) & \hfil 0.000310(9) & \hfil 0.0016(9)\\\cline{2-8}
                &  \hfil $Q_A\land Q_C$ & \hfil 2.0 mm & \hfil 22 & \hfil 41,277 & \hfil 0.0005(1) & \hfil 0.000204(6) & \hfil 0.0010(7)\\\cline{2-8}
                &  \hfil $Q_A\land Q_B\land Q_C$ & \hfil - & \hfil 1 & \hfil 15,389 & \hfil 0.00006(6) & \hfil 0.00000274(9) & \hfil 0.0004(6)\\\cline{2-8}
 \hline
\end{tabular}
\caption{\textbf{Summary of observed, background, and extracted rates.} Observed number of switches and total measurement time leading to observed switching rates $r_i^{\rm obs}$, expected random background coincidence rates $r_i^{\rm background}$, and extracted poisoning event rates $r_i$ for each qubit and qubit combination across both chips for 40 moving averages and a window size of 40, corresponding to the rates plotted in Fig.~4(b) of the main paper.}
\label{observed_rates_and_probs_table}
\end{table*}

\section{Supplementary Note 12: QP Parity Switching Rates on Different Cooldowns} \label{sec:cooldowns}

\renewcommand{\sectionautorefname}{7}

As described in Supplementary Note \autoref{PSD-sec}\hspace*{-1mm}, an unplanned power outage caused our experiment to be split between two cooldowns. Most of the data was collected during the first cooldown, after the dilution refrigerator had been cold for several months. Data measured during the second cooldown was taken within a few weeks of the start of the cooldown. Supplementary Table~\ref{observed_parity_rate_table_nonCu_afterOutage} compares the observed parity switching rates and extracted poisoning rates for the non-Cu chip on the two cooldowns. Supplementary Table~\ref{observed_parity_rate_table_Cu_afterOutage} makes the same comparison for the Cu chip. Although the cryostat was not opened in between the cooldowns and nothing was changed in the wiring, filtering, or shielding, the shorter time period after the start of the second cooldown likely resulted in incomplete thermalization of the radiative environment of the qubit, potentially involving amorphous, non-metallic elements in some of the microwave components or qubit packaging that could slowly release heat over long timescales. This would lead to higher effective blackbody temperatures and a larger flux of THz photons or enhancements to other sources of pair-breaking phonons, such as heat-only events \cite{sAastrom2006, sArmengaud2016,sAnthony2022}, thus resulting in the higher poisoning rates observed on both chips. We note that Ref.~\cite{sMannila2022} also reported a slow decay in QP poisoning rates of a mesoscopic superconducting island over a timescale of several weeks with no clear mechanism for the source of the poisoning.

\begin{table*}[!htbp]
\begin{tabular}{ |p{1.65cm}||p{1.32cm}|p{1.32cm}|p{1.32cm}||p{1.5cm}|p{1.5cm}|p{1.5cm}||p{2.2cm}|  }
 \hline
 \multicolumn{8}{|c|}{Observed parity switching rates (s$^{-1}$) for the non-Cu chip} \\
 \hline
 \hfil {\bf Cooldown}& \hfil $Q_A$ & \hfil $Q_B$ & \hfil $Q_C$ & \hfil $Q_A\land Q_B$ & \hfil $Q_B\land Q_C$ & \hfil $Q_A\land Q_C$ & \hfil $Q_A\land Q_B\land Q_C$\\
 \hline
 \hfil 1 & \hfil 0.299(3)  & \hfil 0.301(4) & \hfil 0.252(3) & \hfil 0.065(2) & \hfil 0.060(2) & \hfil 0.057(2) & \hfil 0.012(1)\\
  \hline
 \hfil 2 & \hfil 0.505(5)  & \hfil 0.508(4) & \hfil 0.495(8) & \hfil 0.170(4) & \hfil 0.161(6) & \hfil 0.162(8) & \hfil 0.042(5)\\
 \hline
 \multicolumn{8}{|c|}{Extracted poisoning event rates 
 (s$^{-1}$)} \\
 \hline
 \hfil 1 & \hfil 0.20(2)  & \hfil 0.19(2) & \hfil 0.12(2) & \hfil 0.18(1) & \hfil 0.17(1) & \hfil 0.15(1) & \hfil 0.064(9)\\
  \hline
 \hfil 2 & \hfil 0.01(4)  & \hfil 0.02(4) & \hfil 0.03(5) & \hfil 0.35(3) & \hfil 0.31(3) & \hfil 0.32(4) & \hfil 0.33(3) \\
 \hline
 \hfil {\it spacing} & \hfil - & \hfil - & \hfil - & \hfil 5.3 mm & \hfil 4.5 mm & \hfil 2.0 mm & \hfil -\\
 \hline
\end{tabular}
\caption{\textbf{Comparison of rates for non-Cu chip between cooldowns.} Observed parity switching rates and extracted poisoning event rates for non-Cu chip on the first and second cooldowns with no poisoning from injector junction. 
}
\label{observed_parity_rate_table_nonCu_afterOutage}
\end{table*}

\begin{table*}[!htbp]
\begin{tabular}{ |p{1.65cm}||p{1.32cm}|p{1.32cm}|p{1.32cm}||p{1.5cm}|p{1.5cm}|p{1.5cm}||p{2.2cm}|  }
 \hline
 \multicolumn{8}{|c|}{Observed parity switching rates (s$^{-1}$) for the Cu chip} \\
 \hline
 \hfil Cooldown& \hfil $Q_A$ & \hfil $Q_B$ & \hfil $Q_C$ & \hfil $Q_A\land Q_B$ & \hfil $Q_B\land Q_C$ & \hfil $Q_A\land Q_C$ & \hfil $Q_A\land Q_B\land Q_C$\\
 \hline
 \hfil 1 & \hfil 0.0221(3) & \hfil 0.0336(5) & \hfil 0.0230(5) & \hfil 0.0008(1) & \hfil 0.0008(2) & \hfil 0.0005(1) & \hfil 0.00006(6)\\
  \hline
 \hfil 2 & \hfil 0.056(2) & \hfil 0.053(2) & \hfil 0.039(1) & \hfil 0.0051(9) & \hfil 0.005(1) & \hfil 0.0047(7) & \hfil 0.0003(3)\\
 \hline
 \multicolumn{8}{|c|}{Extracted poisoning event rates 
 (s$^{-1}$)} \\
 \hline
 \hfil 1 & \hfil 0.041(1)  & \hfil 0.063(1) & \hfil 0.043(2) & \hfil 0.0019(7) & \hfil 0.0016(9) & \hfil 0.0010(7) & \hfil 0.0004(6)\\
  \hline
 \hfil 2 & \hfil 0.082(8) & \hfil 0.07(1) & \hfil 0.047(8) & \hfil 0.015(6) & \hfil 0.017(6) & \hfil 0.016(5) & \hfil 0.000(3)*\\
 \hline
 \hfil {\it spacing} & \hfil - & \hfil - & \hfil - & \hfil 5.3 mm & \hfil 4.5 mm & \hfil 2.0 mm & \hfil -\\
 \hline
\end{tabular}
\caption{\textbf{Comparison of rates for Cu chip between cooldowns.} Observed parity switching rates and extracted poisoning event rates for Cu chip on the first and second cooldowns with no poisoning from injector junction. 
*For the Cu chip on the second cooldown, the solution to the system of equations in Supplementary Eq.~(\ref{prob_eqs}) results in a small negative value for the three-fold coincidence poisoning event rate that is consistent with zero based on the calculated uncertainty.
}
\label{observed_parity_rate_table_Cu_afterOutage}
\end{table*}

\section{Supplementary Note 13: Offset Charge Measurements}

\begin{figure}[htbp!]
\centering
 \includegraphics[width=6.8in]{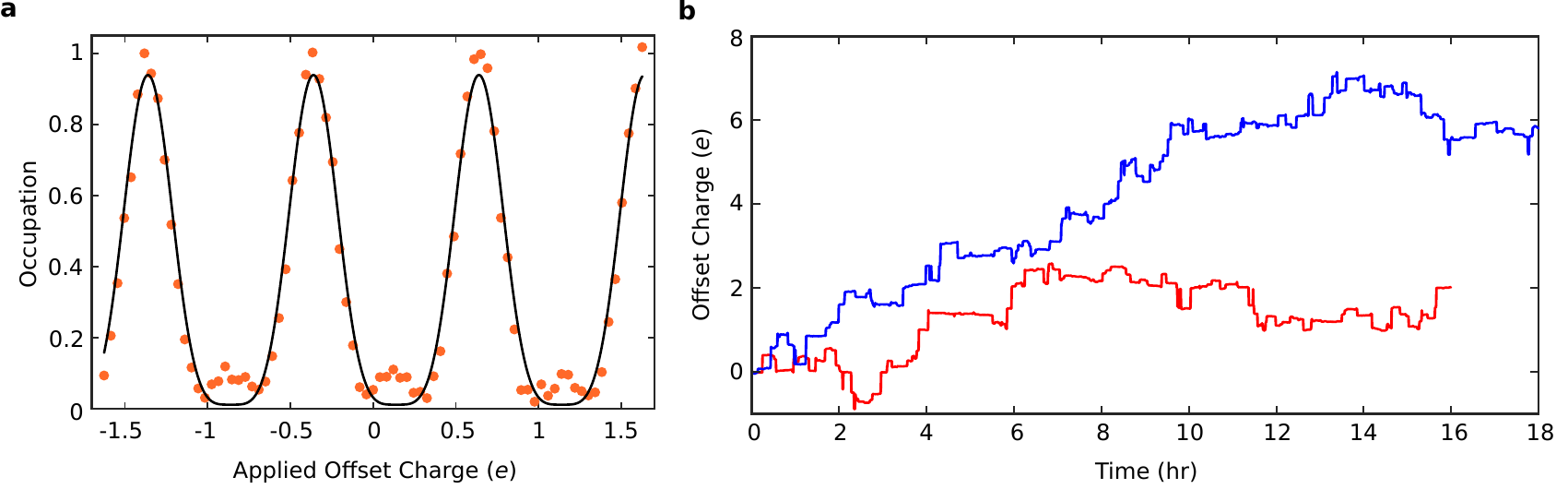}
  \caption{{\bf Offset charge measurements.} (a) Example charge tomography measurement (orange) and fit to Supplementary Eq.~(\ref{1stateprob})  (black) for $Q_B$ on the non-Cu chip. (b) Offset charge vs. time for $Q_B$ on the non-Cu chip (red) and $Q_A$ on the Cu chip (blue).
  \label{fig:CTomo}}
\end{figure}

For a charge-sensitive qubit, besides the parity-mapping sequence, one can also perform a charge tomography sequence to measure the environmental offset charge, provided the qubit has a charge-bias line \cite{sChristensen2019,sWilen2021}. The Ramsey sequence involves two $X/2$ pulses with an idle time $t_i=1/2\delta f$, where $2\delta f$ is the maximum charge dispersion for the qubit. A qubit measurement at the end of the sequence results in a 1-state probability:
\begin{equation}
    P_1=\frac{1}{2}\left[d+\nu\cos\left(\pi\cos 2\pi n_g \right) \right],
\label{1stateprob}
\end{equation}
where $n_g$ is the sum of the externally applied gate charge $n_g^{\rm ext}$ and the environmental offset charge $\delta n_g$; $d$ and $\nu$ are fitting parameters. Supplementary Fig.~\ref{fig:CTomo}(a) shows an example charge tomography trace for $Q_B$ on the non-Cu chip and a fit to Supplementary Eq.~(\ref{1stateprob}).

The charge tomography measurement sequence takes 28~(28.8)~s and we repeat this sequence 2000 (2250) times for $Q_B$ ($Q_A$) on the non-Cu (Cu) chip. ($Q_B$ on the non-Cu chip; $Q_A$ on the Cu chip). From the fit to each tomography scan, we extract $\delta n_g$, which we plot as a function of time over 16 (18) hours for the non-Cu (Cu) chips [Supplementary Fig.~\ref{fig:CTomo}(b)]. 
From these traces, we find that large charge jumps ($\Delta q >0.1 e$) occur at a rate of 0.0012(1)~s$^{-1}$ and 0.0011(1)~s$^{-1}$ for a qubit on the Cu and non-Cu chips, respectively.

Based on this rate of offset charge jumps, we can estimate the rate of $\gamma$ impacts on the chip $R_{\gamma}$ by following the detailed analysis in Ref.~\cite{sWilen2021}. In this case, the authors obtained $R_{\gamma}=0.0198$~s$^{-1}$ from similar measurements of offset charge jump rates, combined with detailed modeling of the effective charge sensing area of their qubits (19,902~$\mu$m$^2$) and simulations of the charge dynamics in the Si substrate. We can approximate the charge sensing area for our qubits by taking this to be the area of the qubit shunt capacitor island extended out to half of the distance between the island and ground plane pocket (6612~$\mu$m$^2$). To estimate $R_{\gamma}$ for our experiment, we scale the corresponding value in Ref.~\cite{sWilen2021} by the 
ratio of the charge sensing area in Ref.~\cite{sWilen2021} to that for our qubit, the ratio of our measured offset charge jump rate to that in Ref.~\cite{sWilen2021} (0.00135~s$^{-1}$), and the ratio of our qubit chip area [$(8\,{\rm mm})^2$] to that in Ref.~\cite{sWilen2021} [$(6.25\,{\rm mm})^2$], leading to the estimate $R_{\gamma}=0.083(8)$~s$^{-1}$ in our system.

\section{Supplementary Note 14: Editing of Device Images} \label{sec:image_editing}

\begin{figure}[htbp!]
\centering
 \includegraphics[width=6.8in]{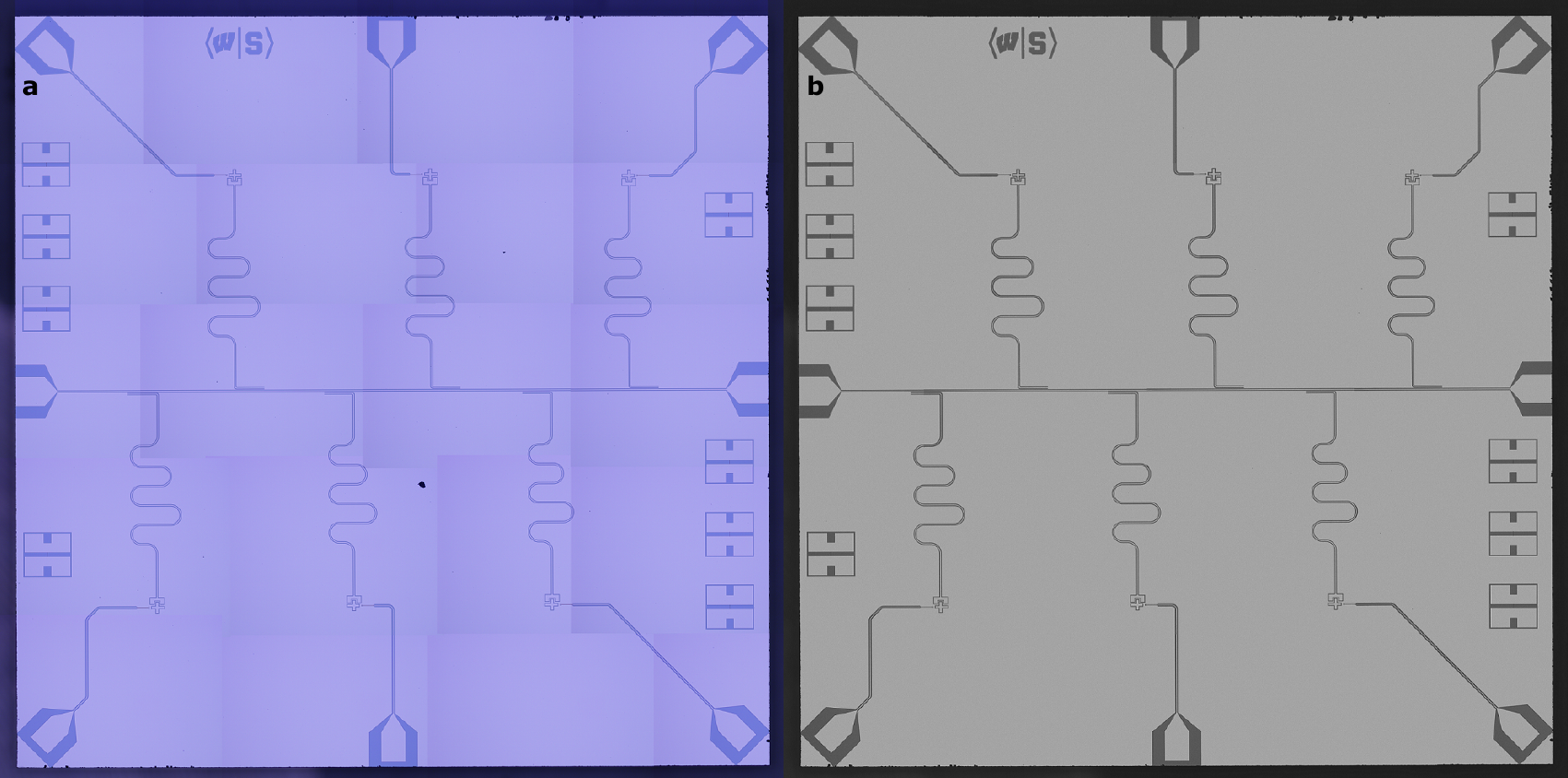}
  \caption{{\bf Raw device image editing.} (a) Tiled layout of the 20 raw images stitched together to make the full chip image. (b) Final chip image after editing.
  \label{fig:imageEditing}}
\end{figure}

The image presented in Supplementary Fig.~\ref{fig:imageEditing}(a) was made by stitching together 20 optical micrographs to achieve a full chip picture. Once each micrograph was aligned, the composite image was converted to grayscale and the contrast increased. At this step, minor surface contamination was removed digitally to limit distraction from important device features. Finally, to remove the vignetting present in each individual picture, the image was processed using MATLAB, which identified the range of pixel values for the Nb background and altered each pixel to reflect the average value with some random noise. The result can be seen in Supplementary Fig.~\ref{fig:imageEditing}(b). The images presented in Fig.~1 and Supplementary Fig.~\ref{fig:chipImages} have been false-colored to highlight different parts of the chip.


\end{document}